%% file: mainpaper.tex
\documentclass[11pt,aps,onecolumn,eprint,prc,nofootinbib,noshowkeys,superscriptaddress,floatfix]{revtex4-1}
\usepackage{amsmath,amsfonts,amsthm,amssymb}
\usepackage[dvips]{graphics,graphicx}
\usepackage[usenames,dvipsnames]{color}
\definecolor{darkblue}{RGB}{0,0,196}
\definecolor{darkgreen}{RGB}{0,120,0}
\usepackage[colorlinks=true,linktocpage=true,linkcolor=darkblue,citecolor=red,urlcolor=darkblue]{hyperref}
\usepackage{cancel}
\usepackage{multirow}
\usepackage{longtable}
\usepackage{color}
\usepackage[normalem]{ulem}
\usepackage{hyperref}
\usepackage{bigints}
\usepackage{xparse}
\usepackage{physics}
\usepackage{verbatim}
\usepackage{minibox}
\usepackage{comment}
\usepackage{xcolor}
\usepackage{appendix}
\usepackage{mathtools}
\usepackage{slashed}
\usepackage{subfigure}

\input{Command}

\begin{document}


\title{Virial theorem for rigidly rotating matter}



\author{Sourav Dey}
\email{sourav.dey@niser.ac.in}
\affiliation{School of Physical Sciences, National Institute of Science Education and Research, An OCC of Homi Bhabha Nuclear Institute, Jatni-752050, India.}

\begin{abstract}
The scaling property of the thermodynamic free energy (\(\Phi\)) of a system at global equilibrium has been examined using a real-time method known as the virial theorem. We demonstrate these scaling properties through a derived relation based on the general structure of equal-time commutators among Poincaré charges and their densities. This relation is applicable to any renormalizable fields with spin \(\leq 1\), excluding gauge fields. In this particular study, we investigate a rigidly rotating solution (\(\Omega = \text{const}\)) at global equilibrium for massless fermionic matter. It has been shown that the applicability of a hydrodynamic description requires a hierarchy \(\Omega R \gg \Omega \beta_{0}\), where \(R\) is the radius of the cylindrical-shaped rotating matter and \(\beta_{0} = 1/T_{0}\) is the inverse temperature on the rotation axis. Consequently, the thermodynamic free energy \(\Phi\) depends on the angular velocity through the product \(\Omega R\), a dependency that extends to other thermodynamic variables as well. These findings are consistent with recent lattice QCD simulation results. Furthermore, we compute the moment of inertia for massless fermions and estimate the light quark contribution to the total moment of inertia of the Quark-Gluon Plasma (QGP) produced in heavy-ion collisions.
\end{abstract}

\maketitle

\section{Introduction}
\label{sec:Intro}
 The study of rotational phenomena in many-body systems has profound implications across diverse branches of physics~\cite{Watts:2016uzu,Grenier:2015pya,Basar:2013iaa,PhysRevLett.122.177202,Landsteiner:2013sja,
PhysRev.6.239,PhysRev.10.7,de2018superconductivity,PhysRevLett.122.177202, PhysRevResearch.2.033013,Fukushima:2018grm,Becattini:2021lfq}. In recent years, relativistic heavy-ion collision experiments at the RHIC and the LHC have emerged as a unique platform for probing quantum matter under extreme rotation~\cite{PhysRevLett.122.177202, PhysRevResearch.2.033013, Fukushima:2018grm, Becattini:2021lfq}. These experiments reveal that the quark-gluon plasma (QGP) formed in off-central collisions acquires the largest vorticity, with angular velocities reaching \( \Omega \sim 10^{22} \, \text{s}^{-1} \, (\approx 7 \, \text{MeV}) \)---the highest vorticity ever observed in a laboratory setting \cite{STAR:2017ckg,Becattini:2020ngo,Huang:2020dtn,Wang:2017jpl}. This vorticity manifests directly in the spin polarization of hyperons (e.g., \( \Lambda \), \( \bar{\Lambda} \)), where the alignment of particle spins with the fluid's rotational structure\footnote{ For the sake of simplicity one may consider the structure to be analogous to a rigidly rotating cylinder.}~\cite{Betz:2007kg} successfully explains experimental polarization data~\cite{ Becattini:2021lfq,STAR:2017ckg, STAR:2007ccu}, linking spin and rotation in a way similar to the Barnett effects observed in its non-relativistic counterpart~\cite{PhysRev.6.239,PhysRev.10.7}. Measurement of the global polarization of hyperons shows good agreement with hydrodynamic prediction \cite{Florkowski:2018fap,Becattini:2022zvf,Chen:2015hfc,Zhang:2020hha,Florkowski:2017ruc}, which implies that polarization phenomena are driven by the collective properties of the system rather than by a specific hadron-dependent coupling. Given the effectiveness of the hydrodynamic description for the observables of heavy ion collision (HIC) in earlier studies \cite{Heinz:2013th,Romatschke:2007mq,Romatschke:2017ejr}, the inclusion of the rotation is a natural extension to understand the local properties of QGP in terms of local vortical structure \cite{Jiang:2016wvv,Becattini:2007sr,Csernai:2013bqa,Deng:2016gyh,Jiang:2016woz}.

In particular, the aspect of thermodynamic systems under rotation within the grand canonical ensemble formalism captures the effect of rotation on the conserved charges or thermodynamic properties \cite{Vilenkin:1978hb,Vilenkin:1980ft,Vilenkin:1980zv}. Consequently, the impact of rotation on the equation of state reveals new features of the QCD phase structure \cite{Chernodub:2020qah,Braga:2023qej,Zhao:2022uxc,Chen:2022smf,Chernodub:2016kxh,PhysRevD.102.114023,PhysRevD.93.104052,PhysRevLett.117.192302,PhysRevD.95.096006,Chen:2020ath,PhysRevD.107.114502}. The prevailing theoretical frameworks \cite{Chen:2015hfc,Sadooghi:2021upd,Wang:2018sur,Yamamoto:2013zwa,Braguta:2023yjn} consider a simple rigidly rotating thermal system, where each physical point in the matter rotates with the same angular velocity, $\Omega$ around a fixed axis. Although the expanding fireball (along the beam axis) is not a rigid rotating system, the model of uniform rotating global equilibrium provides an understanding of global spin polarization of produced hadrons \cite{Becattini:2020ngo,Wang:2017jpl}. However, with the same simple model, there is a lack of consensus between the analytical and numerical results. For example, in several investigations \cite{Huang:2020dtn, Fujimoto:2021xix}, the effect of rotation is independent of the size of the system $R$ and depends solely on some function of $\Omega$. In some recent lattice calculations \cite{PhysRevD.103.094515,Braguta:2020biu}, it has been found that the effect of rotation on thermodynamic properties as a function of the angular velocity comes through the common product $\Omega R$.

Specifically, the interplay between rotation and the finite size of the system introduces a natural causal boundary at a distance \( \Omega^{-1} \), requiring that any physical description satisfies the constraint \( \Omega R \leq 1 \), where \( R \) denotes the radial extent of the system. On the other hand, for a hydrodynamic description to remain valid, the system must satisfy $R \gg \lambda_{th} $ \cite{Bhattacharyya:2007vs}, where $\lambda_{th}$ is the thermal de-Broglie wavelength. This constraint ensures that collective behaviour dominates over microscopic thermal fluctuations, maintaining the validity of the hydrodynamic formalism \footnote{These constraints dictate the regimes in which a quasiparticle-based description remains valid and provide a fundamental limit on the role of rotation in thermal QCD systems.}. In this work, the dependence of the thermodynamic variables on system's size, similar to Refs.~\cite{PhysRevD.102.114023, PhysRevD.93.104052,PhysRevLett.117.192302,Chernodub:2016kxh,Braguta:2023tqz,Braguta:2020biu,Braguta:2023iyx,PhysRevD.107.114502,Braguta:2020biu}, has been motivated by the extensivity properties of the thermodynamic potential.

We use the formalism of statistical mechanics with the grand canonical density operator $\hat{\rho}$ to study the thermodynamic properties of a system at global equilibrium with the Poincare charges $\hat{P}^{\mu},\hat{J}^{\mu\nu}$ as well as some underlying matter charges $\hat{Q}_{a}$ \cite{Vilenkin:1978hb,Vilenkin:1980ft,Vilenkin:1980zv}(also see Refs. \cite{Becattini:2009wh,Becattini:2011ev,Becattini:2012pp,Becattini:2012tc,Becattini:2014yxa,Becattini:2015nva,Becattini:2016stj,Becattini:2017ljh,Buzzegoli:2023yut,Becattini:2018duy,Buzzegoli:2018wpy,Becattini:2019poj,Becattini:2021lfq,Chernodub:2016kxh,Chernodub:2020qah,Florkowski:2017ruc} for recent developments). In thermodynamics, the description of a closed system is made by introducing extensive variables (e.g. volume $V$, total energy $E$, entropy $S$, etc.) to capture the large-scale properties of the matter, irrespective of the specific small-scale physics of the system. 
At equilibrium, the extensivity property of the logarithm of the partition function $\text{log}Z$ (or equivalently, the thermodynamic potential) can be ensured by imposing the scaling property $\text{log}Z\rightarrow\lambda^{3}\text{log}Z$ under $\mxb\rightarrow\lambda\mxb$, which also gives the thermodynamic consistency conditions \cite{Kovtun:2022vas,Kovtun:2019hdm, Banerjee:2012iz,Bhattacharyya:2007vs}. Interestingly, a connection between this extensivity property and the virial theorem was established for both classical and quantum systems \cite{landau1984statistical,Huang1987StatisticalMechanics,Landau_Physical_kinetics,green1948general,born1947general,marc2007virial}, by equating the hydrodynamic pressure (defined via a space-like trace over ensemble averaged energy-momentum tensor) with the thermodynamic pressure (defined via a derivative of thermodynamic potential with respect to volume). As a consequence of extensivity property of the thermodynamic potential, the intensivity of the thermodynamic pressure is guaranteed.

In this article, we present a field-theoretic version of the virial theorem  \cite{Lin:2015cia,Toyoda:1998ub,Ordonez:2015vaa} by invoking a spatial scaling transformation, in the line of Ref. \cite{Landsman:1986uw}.
Specifically, we apply this framework for rotating thermal systems, considering massless fermions to represent the light-quark contributions of the rotating QGP in a qualitative manner. The effect of rotation on thermodynamic pressure is expressed as a function of angular velocity $\Omega$. Due to the intensive nature of thermodynamic pressure, the $\Omega$-dependency appears through dimensionless combinations $\Omega R$ and $\Omega\beta_{0}$, where $\beta_{0}=\frac{1}{T_{0}}$ is the inverse of the temperature on the rotation axis. Recalling the stringent condition $R\gg\lambda_{th}$~\cite{Bhattacharyya:2007vs,Kovtun:2022vas,Liu:2018kfw}, where the thermal wavelength goes as inverse temperature, i.e., $\lambda_{th}\sim\frac{1}{T_{0}}$ \cite{Bhattacharyya:2007vs,Dey:2024crk} for a massless (or strongly coupled) system, we show that a separation in magnitude arises between these two dimensionless quantities, i.e.,  $\Omega R\gg\Omega\beta_{0}$. Consequently, the system size dependency of thermodynamic properties is pronounced through the domination of the common product $\Omega R$.

This article is organised as follows: In Sec.~\ref{Sec1}, a general description of the finite temperature grand canonical system with Poincare charges and matter charges has been discussed, and the definition of the partition function has been reviewed.
A field-theoretic counterpart of the viral theorem has been reviewed through the introduction of the rigid scaling transformation and its action on the local operator in Sec.~\ref{SecVirial}. We also show that the general structure of equal-time commutator (ETC) among different components of energy-momentum tensors, as well as the charge currents, leads to a general relation \rf{GenP} without considering any particular renormalized matter fields of spin $\leq1$ (except Gauge theories). Sec.~\ref{Sec4} is devoted to understanding this general relation with two specific solutions of the Killing equation at the global equilibrium, namely nonrotating homogeneous and rotating inhomogeneous solution. We show that the imposition of the extensivity of the thermodynamic potential gives us an insight into the effect of rotation on thermodynamic quantities such as the moment of inertia. We further find for the noninteracting fermionic system, which turns out to be positive and meets the criterion of thermodynamical stability. An estimation of the moment of inertia for the quark content in the QGP has been done in Sec.~\ref{conclusion} by considering appropriate values of physical parameters with a simple model of rigidly rotating massless fermions, and shows the dominant nature of the common product $\Omega R$, which is similar to the observation of Refs.~\cite{PhysRevD.103.094515,PhysRevD.107.114502,Braguta:2020biu,Chernodub:2016kxh} in the slow rotation limit.

\textit{Notation and convention:} Throughout the paper, we use the metric tensor that has the signature $(+,-,-,-)$. Three-vectors are denoted by bold fonts (e.g. $\mathbf{a}$) as well as the usual forms (e.g. $\vec{a}$). The scalar products for both three- and four-vectors are denoted by a dot, i.e., $a \cdot b = a^0 b^0 - \av \cdot \bv$ (or $=a^{0} b^{0} - \vec{a} \cdot \vec{b}$).

\section{Finite temperature QFT with global charges}\label{Sec1}
In Minkowski space, the grand canonical density operator at global equilibrium is expressed as  
\begin{align}
    \hat{\rho} = \frac{1}{Z} \exp\left[-b_{\mu} \hat{P}^{\mu} + \frac{1}{2} \omega_{\mu\nu} \hat{J}^{\mu\nu} + \sum_{a} \alpha_{a} \hat{Q}_{a} \right],\label{rhoden}
\end{align}  
where \(\hat{P}^{\mu}\), \(\hat{J}^{\mu\nu}\) represents the ten conserved mechanical integrals of motion: the momentum four vector and angular momentum, respectively. Here, the operator \( \hat{Q}_{a} \) denotes additional conserved charges, with different charges labeled by the index \( a \). The Lagrange multipliers \(b_{\mu}\), \(\omega_{\mu\nu}\), and \(\alpha_{a}\) are associated with these conserved quantities. The first two constitute the Killing field $\beta_{\mu}(x)$ as
\begin{align}
    \beta_{\mu}(x)=b_{\mu}+\omega_{\mu\nu}x^{\nu}\,,\label{Kfi}
\end{align}
and $\alpha_{a}(=\text{const})$ having the homogeneous profile are the solution of the global equilibrium conditions
\begin{align}
    \partial_{\mu}\beta_{\nu}(x)+\partial_{\nu}\beta_{\mu}(x)=0\,\,\text{and}\,\, \partial_{\mu}\alpha_{a}=0\,,\label{KillCon}
\end{align}
correspondingly. Here, $\beta_{\mu}(x)$ and $\alpha_{a}$ help define the proper comoving temperature $T(x)=\big(\sqrt{\beta_{\mu}\beta^{\mu}}\big)^{-1}$ and the comoving chemical potential $\mu_{a}(x)=\alpha_{a} T(x)$, respectively.

Here, the partition function,
\begin{align}
    Z \!=\! \operatorname{Tr} \!\left[\exp\left(\!-b_{\mu} \hat{P}^{\mu} + \frac{1}{2} \omega_{\mu\nu} \hat{J}^{\mu\nu} +\! \sum_{a} \alpha_{a} \hat{Q}_{a} \right)\right]\label{PartN}
\end{align} 
is defined for a finite system size, ensuring proper normalization of the density operator. Moreover, the extensive nature of $\text{log}(Z)$ helps to define the thermodynamic pressure (also called the kinetic pressure)~\cite{Landsman:1986uw,marc2007virial}. With the assistance of this density operator, one can define the ensemble average of an operator $\hat{O}(x)$ as
\begin{align}
    \Big\langle\hat{O}(x)\Big\rangle=\text{Tr}\Big(\hat{\rho}\,\hat{O}(x)\Big)\,,\label{AvO}
\end{align}
where the trace is defined for the finite volume.

\section{Virial theorem }
\label{SecVirial}
For both classical and quantum systems in equilibrium, the virial theorem \cite{Landsman:1986uw,marc2007virial,Toyoda:1998ub,Ordonez:2015vaa} establishes an equivalence between the hydrostatic pressure, defined by the space-like trace of the energy-momentum tensor, and the thermodynamic pressure, derived from the partition function. The field-theoretic version of this theorem has been discussed in Ref.~\cite{Landsman:1986uw}, and will be employed in this section to explain the system in global equilibrium with the density operator from \rf{rhoden}.
\subsection{Dilatation as pseudogauge transformation}
\label{DilPSG}
In the context of field theory, the virial theorem can be proved by involving a spatial dilatation, which scales the space-time  coordinate and fields as
\begin{align}
    t\rightarrow t,\,\mathbf{x}\rightarrow e^{\lambda}\mathbf{x},\,\Phi(x)\rightarrow e^{\Delta\lambda}\Phi(t,e^{\lambda}\mathbf{x}),
\end{align}
where $\Delta$ is the scale dimension of the field considered. The infinitesimal form of this transformation gives
\begin{align}
\mathbf{x}\rightarrow(1+\lambda)\mathbf{x},\,\Phi(x)\rightarrow\big(1+\lambda(\Delta+\mathbf{x}\cdot\partial)\big)\Phi(x)\,.
\end{align}
So, the Noether charge \cite{francesco2012conformal}, which generates the spatial dilatation, can be written as
\begin{align}
    \hat{D}(t)=\int d^{3}\mathbf{x}\Big[x_{i}\hat{T}_{B}^{0 i}(x)+\hat{V}^{0}(x)\Big]\,.\label{DilC}
\end{align}
Here, $\hat{T}^{\mu\nu}_{B}$ is the Belinfante energy-momentum tensor, and $\hat{V}^{0}$ is the time component of the virial field which vanishes for spin $\frac{1}{2}$ field and the renormalizable massive spin $1$ field. The virial field has a nontrivial structure for the scalar and gauge theories. However, in cases other than gauge theories \footnote{In gauge theories, the virial field $\hat{V}^{\mu}$ depends on the choice of gauge. However, this gauge dependence does not survive when taking the grand canonical expectation over physical states \cite{Landsman:1986uw}. These subtleties are not explored in this article; a generalization of the analysis is left for future work.} the dilatation generator \rf{DilC} can be re-expressed  for all renormalizable theories as 
\begin{align}
    \hat{D}(t)=\int d^{3}\mathbf{x}\,x_{i}\hat{T}^{0i}(x)\,,\label{DiLI}
\end{align}
in terms of the ``improved" energy momentum tensor \cite{Callan:1970ze,Coleman:1970je,PhysRevD.2.1541}, which is defined as 
\begin{align}
    \hat{T}^{\mu\nu}=\hat{T}^{\mu\nu}_{B}-\frac{1}{6}\sum_{a}\Big(\partial^{\mu}\partial^{\nu}-g^{\mu\nu}\partial^{2}\Big)\phi^{2}_{a}\,\label{ITV}.
\end{align}
In the previous equation, the difference between the Belinfante and the improved energy-momentum tensor occurs only for the scalar fields $(\phi_{a})$.

As mentioned in the previous section, we assume that the fluid enjoys time, space, Lorentz (boost +  rotation) invariance as well as global $U(1)_{a}$ symmetry invariance generated by the corresponding charges,  $\hat{P}^{0}$\,(or Hamiltonian $\hat{H}$), $\hat{P}^{i}$, $\hat{J}^{\mu\nu}$\,($\HJ^{0i}+\HJ^{ij})$, and $\hat{Q}_{a}$, respectively.
We adopt a static coordinate system $(x^{0}, x^{i})$ or $(t,\mathbf{x})$, namely the laboratory (LAB) frame that possesses the Killing vector $\partial_{0}$. Relative to this static coordinate the fluid flows with four velocity $u_{\mu}(x)=\frac{1}{\sqrt{\beta^{2}}}\beta_{\mu}(x)$, where $\beta_{\mu}(x)$ defines a timelike Killing vector that helps to define thermodynamics consistently. Thus, in the static frame, the spacetime symmetry generators are these hermitian operators $\hat{P}_{\mu}=i\partial_{\mu},\,$ and $\hat{J}_{\mu\nu}=i\Big(x_{\mu}\partial_{\nu}-x_{\nu}\partial_{\mu}\Big)$ corresponding to the spacetime translation and the Lorentz transformation in differential representation, respectively. On the other hand, the diffeomorphism and global $U(1)$ invariance of the action relate these generators to the tensor densities\footnote{Here, we adopt the Belinfante pseudo-gauge, in which the total angular momentum \(J^{\mu\nu}\) receives contributions solely from the orbital part due to the vanishing spin current. Subsequently, in Eqs.~\eqref{ImTP} and \eqref{ImTJ}, the conserved charges are defined using the improved energy--momentum tensor \(\hat{T}^{\mu\nu}(x)\).}
\begin{align}
    \hat{P}^{\mu}&=\int d^{3}\mathbf{x}\,\hat{T}_{B}^{0\mu}(x)\,,\label{PVB}\\
    \hat{J}^{\mu\nu}&=\int d^{3}\mathbf{x}\,\hat{J}_{B}^{0\mu\nu}(x)\nonumber\\
    &=\int d^{3}\mathbf{x}\,\Big(x^{\mu}\hat{T}^{0\nu}_{B}(x)-x^{\nu}\hat{T}^{0\mu}_{B}(x)\Big)\,,\label{JVB}\\
    \hat{Q}_{a}&=\int d^{3}\mathbf{x}\,\hat{J}_{a}^{0}(x)\,.\label{Qa}
\end{align}
Here, the angular momentum tensor density is defined by $\hat{J}^{\mu\alpha\beta}_{B}=\Big(x^{\alpha}\hat{T}_{B}^{\mu\beta}-x^{\beta}\hat{T}_{B}^{\mu\alpha}\Big)$, and $\hat{J}^{\mu}_{a}$ is the conserved current density associated with the $U(1)_{a}$ symmetry.
While, the global Poincare charges in \rf{PVB} and \eqref{JVB} are defined in terms of the Belinfante energy-momentum tensor $\hat{T}^{\mu\nu}_{B}(x)$, they can be re-expressed with some new energy-momentum tensor $\hat{T}^{\prime\mu\nu}(x)$, provided the latter satisfies,
\begin{align}
    \hat{T}^{\prime\mu\nu}(x)&=\hat{T}^{\mu\nu}_{B}(x)+\frac{1}  {2} \partial_\lambda \left( 
\hat{\Phi}^{\lambda, \mu \nu} 
+\hat{\Phi}^{\nu, \mu \lambda}
+\hat{\Phi}^{\mu, \nu \lambda} \right)\nonumber\\
&\,\quad+\frac{1}{2}\partial_{\lambda}\partial_{\rho}\hat{\Psi}^{\lambda\rho\mu\nu}\,.\label{PGTT}
\end{align}
To keep the global angular momentum tensor $\hat{J}^{\mu\nu}$ unaltered, we have to simultaneously modify the angular momentum tensor as
\begin{align}
    \hat{J}^{\prime\mu\alpha\beta}=x^{\alpha}\HT^{\prime\mu\beta}-x^{\beta}\hat{T}^{\prime\mu\alpha}+\HS^{\prime\mu,\alpha\beta}\,,
\end{align}
where,
\begin{align}
\hat{S}^{\prime \lambda, \mu \nu}&=- \hat{\Phi}^{\lambda, \mu \nu} + \partial_\rho \hat{Z}^{\mu\nu, \lambda \rho}\,.\label{PGST}
\end{align}
Here, these additional boundary terms $\hat{\Phi}^{\lambda\mu\nu}$, $\hat{\Psi}^{\lambda\rho\mu\nu}$ and $\hat{Z}^{\mu\nu,\alpha\beta}$ are known as super-potential, they possess the following symmetries with respect to the exchange of the indices
\begin{align}
   \hat{\Phi}^{\lambda, \mu \nu} &= -\Phi^{\lambda, \nu \mu}, \nonumber \\
   \hat{\Psi}^{\lambda\rho\mu\nu}&=\frac{1}{3!}\hat{\Psi}^{[\lambda\rho\mu]\nu}\,,\label{phisym}
\end{align}
and
\begin{align}
  \hat{Z}^{\mu\nu, \lambda \rho} &= - \hat{Z}^{\nu\mu, \lambda \rho}= -\hat{Z}^{\mu\nu, \rho \lambda}\,.\label{Zsym}
\end{align}
These symmetry properties are crucial for ensuring the conservation of the local energy-momentum (through $\partial_{\mu}\HT^{\prime\mu\nu}=0$) and the local angular momentum (through $\partial_{\lambda}\HJ^{\prime\lambda\alpha\beta}=0$), provided that the corresponding conservation laws were satisfied in the Belinfante choice (i.e., $\partial_{\mu}\HT^{\mu\nu}_{B}=0$ and $\partial_{\lambda}\HJ_{B}^{\lambda,\alpha\beta}=0$). Despite all these symmetry constraints of the Eqs.~\eqref{phisym}-\eqref{Zsym} imposed on the indices of the super potentials, there is an arbitrariness in the choice of tensor densities $(\HT^{\prime\mu\nu}(x), \HJ^{\prime\lambda\mu\nu}(x)$) with the same total Poincare charges $\PV^{\mu},\,\JV^{\mu\nu}$, which is known as 
\textit{pseudogauge freedom} in the literature \cite{Belinfante:1940,BELINFANTE1939887,rosenfeld1940tenseur,Hehl:1976vr,Speranza:2020ilk,Dey:2023hft}. Consequently, the improved energy-momentum tensor defined in \rf{ITV} can be interpreted as a pseudogauge transformation as given in \rf{PGTT} and \rf{PGST}, with the choice of $\hat{\Phi}^{\mu,\alpha\beta}(x)=0,\hat{Z}^{\lambda\rho,\alpha\beta}(x)=0$ but nonzero $\hat{\Psi}^{\mu\nu\alpha\beta}(x)$~\cite{Jackiw:1971dp,Callan:1970ze,Speranza:2020ilk}.

The rigid scaling transformation introduces the dilatation $\hat{D}(t)$ operator in Eq.~\eqref{DilC}, which transforms an operator $\hat{O}(t)$ by
\begin{align}
    \hat{O}(t)\rightarrow \hat{O}(t)+i\lambda\big[\hat{D}(t),\hat{O}(t)\big]\,.
\end{align}
To analyse the scaling properties of the thermal system, one needs to understand the action of the dilation on the density matrix $\hat{\rho}$. Thus, it is crucial to know the ETC relations between the dilatation generator $\HD$ and other symmetry generators. Using the differential representation of dilatation generator $\HD=ix^{i}\partial_{i}$, it can be shown the only non-vanishing commutators are,
\begin{align}
\Big[\hat{D}(t),\hat{P}_{i}\Big]&=-i\hat{P}_{i}\,,\label{PDC}\\
\Big[\hat{D}(t),\HJ_{0i}\Big]&=-i\Big(x_{i}\hat{P}_{0}+x_{0}\hat{P}_{i}\Big)\,,\nonumber\\
&=-i\Big(2x_{0}\hat{P}_{i}-\HJ_{0i}\Big)\,.\label{JDC}
\end{align}

Alternatively, these ETCs can be obtained by using representations given in Eqs.~\eqref{DilC}, \eqref{DiLI}, and \eqref{PVB}-\eqref{Qa}, which involve ETC of the tensor fields $\big(\HT^{0\nu}(x),\HJ^{0}(x),\,\,\text{etc.}\big)$. The explicit computation of the ETC between dilatation generator $\hat{D}(t)$ and the current density $\HJ_{a}^{0}(x)$ associated with the $a$-type charge has been performed in Refs.~\cite{Jackiw:1968lmf,Genz:1970gzi,Gross:1967zz,Beg:1970ai,Levin:1971sw}, yielding
\begin{align}
    i\Big[\HD(t),\HJ_{a}^{0}(x)\Big]=\big(3+x^{i}\partial_{i}\big)\HJ_{a}^{0}(x).\label{charD}
\end{align}
This relation shows that the charge density has a scale dimension of three\footnote{The extra term in the \rf{ITV} does not contribute to this commutator, if $\JV^{0}_{a}$ is a matter current or corresponding to global gauge transformations.}.

Furthermore, the ETCs between the dilatation generator and the space-time generators involve the commutators of energy-momentum tensors; these have been discussed in a general form in Refs.~\cite{Banks:1972jf,Schwinger:1963zz,Landsman:1986uw,Deser:1967zzf,Trubatch:1970zr,Jackiw:1968lmf}. In our particular context the following commutators are required \cite{Deser:1967zzf,Trubatch:1970zr}\footnote{These equal-time commutators hold for any energy-momentum tensor $\HT^{\prime\mu\nu}$, which are connected to the Belinfante $\HT^{\mu\nu}_{B}$ via pseudo gauge transformation as given in \rf{PGTT} .}
\begin{align}
    i\Big[\HT^{00}(t,\mathbf{x}),\HT^{0i}(t,\mathbf{y})\Big]&=\Big[\HT^{ji}(t,\mathbf{x})-g^{ji}\HT^{00}(t,\mathbf{y})\Big]\nonumber\\
    &\,\times\partial_{j}\delta^{3}(\mathbf{x}-\mathbf{y})+\hat{\tau}^{00,0i}(x,y)\label{TT1}
\end{align}
\begin{align}
    i\Big[\HT^{0i}(t,\mathbf{x}),\HT^{0j}(t,\mathbf{y})\Big]&=\Big[\HT^{0j}(t,\mathbf{x})\partial^{i}+\HT^{0i}(t,\mathbf{y})\partial^{j}\Big]\nonumber\\
    &\,\times\delta^{3}(\mathbf{x}-\mathbf{y})+\hat{\tau}^{0i,0j}(x,y)\label{TT2}
\end{align}
The term on the right hand side containing $\partial_{i}\delta^{3}(\mathbf{x}-\mathbf{y})$ is essential to ensure the Poincare algebra is generated by $\hat{P}^{\mu}=\int d^{3}\mathbf{x}\,\HT^{0\mu}(x)$ and $\HJ^{\mu\nu}=\int d^{3}\,\mathbf{x}\,\big(x^{\mu}\HT^{0\nu}(x)-x^{\nu}\HT^{0\mu}(x)\big)$. The last term of Eqs.~\eqref{TT1} and \eqref{TT2}, also known as the Schwinger term, contains a higher derivative of spatial delta functions. Hence, the following integrals must vanish:
\begin{align}
    \int d^{3}\mathbf{x}\,\hat{\tau}^{00,0i}(x)=0,\,\int d^{3}\mathbf{x}\,\hat{\tau}^{0i,0j}(x)=0\,,
\end{align}
and
\begin{align}
    \int d^{3}\mathbf{x}\,\mathbf{x}\,\hat{\tau}^{00,0i}(x)=0,\,\int d^{3}\mathbf{x}\,\mathbf{x}\,\hat{\tau}^{0i,0j}(x)=0\,.
\end{align}
Employing these commutator relations for non-gauge theories, one can arrive at 
\begin{align}
    i\Big[\HD(t),\HT^{00}(t,\mathbf{x})\Big]&=\Big(3+x^{i}\partial_{i}\Big)\HT^{00}(x)-\HT^{i}_{i}(x)\,,\label{DT00C}
\end{align}
\begin{align}
   i\Big[\HD(t),\HT^{0i}(t,\mathbf{x})\Big]&=\Big(4+x^{i}\partial_{i}\Big)\HT^{0i}(x)\,,\label{DT0iC}
\end{align}
which shows that the tensor component of the improved energy-momentum tensor $\HT^{00}(x)$ has no definite scale dimension, whereas $\HT^{0i}(x)$ clearly has a scale dimension of four.
This construction will be useful for analyzing the extensive properties of thermodynamical free energy, as described in the next subsection.
\subsection{Scaling at global equilibrium }
\label{subsecGeQ}
Based on the commutators provided in \rf{PDC} and \rf{JDC}, and the definition of the expectation value at global equilibrium from \rf{AvO}, the following identity can be derived:
\begin{align}
    \Big\langle i\big[\HD(t)\,,&-b_{\mu} \hat{P}^{\mu} + \frac{1}{2} \omega_{\mu\nu} \hat{J}^{\mu\nu} + \sum_{a} \alpha_{a} \hat{Q}_{a}\big]\Big\rangle\nonumber\\
    &=-\omega_{0i}\langle\HJ^{0i}\rangle-\Big(b_{i}+2\omega_{i0}x^{0}\Big)\langle\hat{P}^{i}\rangle\,.\label{DPCC}
\end{align}
Concurrently, one can express the Poincaré charges in terms of the improved energy-momentum tensor $\HT^{\mu\nu}(x)$ as follows:
\begin{align}
    \hat{P}^{\mu}&=\int d^{3}\mathbf{x}\,T^{0\mu}(x)\,,\label{ImTP}\\
    \HJ^{\mu\nu}&=\int d^{3}\mathbf{x}\,\Big(x^{\mu}\HT^{0\nu}(x)-x^{\nu}\HT^{0\mu}(x)\Big)\,,\label{ImTJ}
\end{align}
and the matter charges as given in \rf{Qa}. This aids in expressing the left-hand side of \rf{DPCC} in the following form by using Eqs.~\eqref{charD}, \eqref{DT00C} and \rf{DT0iC}
\begin{align}
   &\Big\langle i\big[\HD(t)\,,-b_{\mu} \hat{P}^{\mu} + \frac{1}{2} \omega_{\mu\nu} \hat{J}^{\mu\nu} + \sum_{a} \alpha_{a} \hat{Q}_{a}\big]\Big\rangle\nonumber\\
    &=\int d^{3}\mathbf{x}\bigg\langle-b_{0}\Big(\mathcal{D}(\mxb)\HT^{00}(x)-\HT^{k}_{k}(x)\Big)\nonumber\\
    &-b_{i}\Big(\mathcal{D}(\mxb)\HT^{0i}(x)+\HT^{0i}(x)\Big)+\omega_{0i}\Big(2x^{0}\HT^{0i}(x)\nonumber\\
    &+\big(\mathcal{D}(\mxb)-1\big)\big(x^{0}\HT^{0i}(x)-x^{i}\HT^{00}(x)\big)+x^{i}\HT^{k}_{k}(x)\Big)\nonumber\\
    &+\frac{\omega_{ij}}{2}\Big(\mathcal{D}(\mxb)\big(x^{i}\HT^{0j}(x)-x^{j}\HT^{0i}(x)\big)\Big)\nonumber\\
    &+\alpha_{a}\Big(\mathcal{D}(\mxb)\HJ^{0}_{a}(x)\Big)\bigg\rangle\,.\label{LhS}
\end{align}
Here, the differential operator $\mathcal{D}(\mxb)=\big(3+x^{i}\partial_{i}\big)$ appears due to the action of the dilatation generator on each local operator $\hat{O}(x)$, which helps to write each integral as
\begin{align}
    \int d^{3}\mxb\,\mathcal{D}(\mxb)\hat{O}(x)=\frac{\partial}{\partial\lambda}\Big(\int d^{3}\mxb\,e^{3\lambda}\hat{O}(t,e^{\lambda}\mxb)\Big)\Big|_{\lambda\rightarrow 0}\,.
\end{align}

Hence, the integrands in the last equation appears with $\mathcal{D}(x)$ can be rewritten as 
\begin{align}
    \Bigg\langle&\int d^{3}\mathbf{x}\,\mathcal{D}(\mathbf{x})\Big(-b_{0}\HT^{00}(x)-b_{i}\HT^{0i}(x)+\omega_{0i}\big(x^{0}\HT^{0i}(x)-x^{i}\HT^{00}(x)\big)+\frac{\omega_{ij}}{2}\big(x^{i}\HT^{0j}(x)-x^{j}\HT^{0i}(x)\big)+\alpha_{a}\HJ^{0}_{a}(x)\Big)\Bigg\rangle\nonumber\\
    &=\Bigg\langle\frac{\partial}{\partial\lambda}\Bigg(\int d^{3}\mxb\,e^{3\lambda}\bigg(-b_{\mu}\HT^{0\mu}(t,e^{\lambda}\mxb)+\omega_{0i}\Big(x^{0}\HT^{0i}(t,e^{\lambda}\mxb)-x^{i}e^{\lambda}\HT^{00}(t,e^{\lambda}\mxb)\Big)+\frac{\omega_{ij}}{2}e^{\lambda}\Big(x^{i}\HT^{0j}(t,e^{\lambda}\mxb)\nonumber\\
    &-x^{j}\HT^{0i}(t,e^{\lambda}\mxb)\Big)+\alpha_{a}\HJ^{0}_{a}(t,e^{\lambda}\mxb)\bigg)\Bigg|_{\lambda\rightarrow 0}\Bigg\rangle\nonumber\\
    &=\partial_{\lambda}\log Z(\lambda)\big|_{\lambda\rightarrow 0}\,,
\end{align}
where,
\begin{align}
    \log Z(\lambda)&=\int d^{3}\mathbf{x}\,e^{3\lambda}\bigg(-b_{\mu}\HT^{0\mu}(t,e^{\lambda}\mxb)+\omega_{0i}\Big(x^{0}\HT^{0i}(t,e^{\lambda}\mxb)-x^{i}e^{\lambda}\HT^{00}(t,e^{\lambda}\mxb)\Big)+\frac{\omega_{ij}}{2}e^{\lambda}\Big(x^{i}\HT^{0j}(t,e^{\lambda}\mxb)\nonumber\\
    &-x^{j}\HT^{0i}(t,e^{\lambda}\mxb)\Big)+\alpha_{a}\HJ^{0}_{a}(t,e^{\lambda}\mxb)\bigg)
\end{align}
is the logarithm of the scaled partition function. Here, $Z(\lambda)$ can be obtained from the definition \rf{PartN} by scaling each of the extensive quantities appearing in the argument of $Z$ (for example, $V \rightarrow e^{3\lambda}V$ or $L \rightarrow e^{\lambda}L$).
Consequently, with the help of the expressions in \rf{DPCC} and \rf{LhS}, one shall arrive at
\begin{align}
    &\int d^{3}\mxb\big(b_{0}+\omega_{0i}x^{i}\big)\langle\HT^{k}_{k}(x)\rangle\nonumber\\
    &=-\partial_{\lambda}\text{log}Z(\lambda)\big|_{\lambda\rightarrow0}\,.\label{GenP}
\end{align}
This shows that the scaling property of the thermodynamic partition function is encoded in the general identity given in \rf{GenP}, which holds for a system in global equilibrium possessing the maximal number of Poincaré charges and various matter charges, within renormalizable non-gauge field theories with spin $\leq1$. The implications of this identity, in the context of different solutions to the global Killing condition \rf{KillCon}, are discussed in the next section.
\section{Extensivity property of $\text{log} Z$}
\label{Sec4}
The global equilibrium state in Minkowski space-time possesses ten constants of motion, encoded in the Killing fields given in \rf{Kfi}, corresponding to the Poincaré symmetry group.
We begin by considering a familiar form of the density operator, which can be recovered from \rf{rhoden} by choosing $b_{\mu} = \frac{1}{T_{0}}(1, \mathbf{0}) = \beta_{0} \delta_{\mu}^{0}$ and $\omega_{\mu\nu} = 0$, which is
\begin{align}
    \hat{\rho} = \frac{1}{Z} \exp\left[-\beta_{0}\hat{H}+ \sum_{a} \alpha_{a} \hat{Q}_{a} \right]\,.\label{rhodenT0}
\end{align}
Here, $\hat{H}$ is the Hamiltonian of the equilibrated system, which measures the energy as seen by an observer moving along the Killing vector $\partial_{0}$ in the static coordinate frame. With this Killing vector, we obtain the homogeneous proper temperature $T_{0}=\beta^{-1}_{0}$. Therefore, the ensemble average of any local operator as given in \rf{AvO} becomes position independent, due to the translational invariance of the density operator \rf{rhodenT0}.

Following the right-hand side of the \rf{GenP}, and with homogeneous thermodynamic parameters, we arrive at
\begin{align}
    &\int d^{3}\mathbf{x}\,\beta_{0}\langle\HT^{k}_{k}(x)\rangle=V\beta_{0}\langle\HT^{k}_{k}(x)\rangle\,.
\end{align}
Here, the left-hand side of \rf{GenP} provides insight into the scaling property of the logarithm of the partition function, $\text{log}Z$. Numerous studies \cite{landau1984statistical,Kovtun:2019hdm,Romatschke:2017ejr,PhysRevD.16.1130,Freedman:1976dm} have shown that robustness of extensivity ensures writing $\text{log}Z(\lambda)=\text{log}Z(\beta_{0},\alpha_{a},Ve^{3\lambda})=e^{3\lambda}\text{log}Z$, which leads to the identity
\begin{align}
    &\beta_{0}\langle\HT^{k}_{k}(x)\rangle\nonumber\\
    &=-\frac{1}{V}\partial_{\lambda}\text{log}Z(\beta_{0},\alpha_{a},Ve^{3\lambda})\big|_{\lambda\rightarrow0}\nonumber\\
    &=-3\frac{\partial}{\partial V}\Big(\text{log}Z(\beta_{0},\alpha_{a},V)\Big)\,.\label{HomP}
\end{align}
The last relation allows us to identify the isotropic hydrodynamic pressure defined through the spatial trace of the energy-momentum tensor $\langle\HT^{k}_{k}(x)\rangle=3 P_{0}$ in the rest frame of the fluid, provided a suitable energy-momentum tensor has been chosen. On the other hand, the thermodynamic pressure - an intensive variable-is defined as follows \cite{landau1984statistical,Huang1987StatisticalMechanics}
\begin{align}
  \mathcal{P}(\beta_{0},\alpha_{a})=-\partial_{V}\Phi(\beta_{0},\alpha_{a},V)\,,  
\end{align}
where $\Phi(\beta_{0},\alpha_{a},V)$ is the thermodynamic potential defined through the partition function 
\begin{align}
    \Phi(\beta_{0},\alpha_{a},V)&=-T_{0}\text{log}Z(\beta_{0},\alpha_{a}, V)\,,\nonumber\\
    &=-\mathcal{P}(\beta_{0},\alpha_{a})V
\end{align}
Hence, \rf{HomP} equates hydrodynamic and thermodynamic pressures, $P_{0}=\mathcal{P}(\beta_{0},\alpha_{a})$, as a consequence of the extensivity of the thermodynamic potential; an identity also known as the virial theorem \cite{marc2007virial,Landsman:1986uw}.

We now turn to the global thermodynamics of a system undergoing pure rotation at constant angular velocity $\Omega$ along the $z$-axis, characterized by $b_{\mu}=(\beta_{0}, \mathbf{0})$ and $\omega_{\mu\nu}=\beta_{0}\Omega(g_{\mu x}g_{\nu y}-g_{\mu y}g_{\nu x})$. The corresponding equilibrium density operator is given by 
\begin{align}
     \hat{\rho} = \frac{1}{Z} \exp\left[-\beta_{0}\hat{H}+\beta_{0}\Omega\hat{J}^{xy}+ \sum_{a} \alpha_{a} \hat{Q}_{a} \right]\,.\label{rhodenT0Om}
\end{align}
which has been captured recent interest, as described in Refs.~\cite{Becattini:2021lfq,landau1984statistical,Vilenkin:1980zv,Chen:2015hfc,Zhang:2020hha,Florkowski:2017ruc,Becattini:2020sww}. Moreover, the Killing vector $\beta_{\mu}(x)$ \cite{Becattini:2016stj} does not lead to a homogeneous thermodynamic equilibrium, because temperature and the chemical potential become position dependent. Specifically become a function of the radial coordinate $r=\sqrt{\mxb^{2}+\mathbf{y}^{2}}$, due to the angular velocity along the $z$-axis, can be written as
\begin{align}
    T(x)=\Gamma(r)T_{0}\,,\,\mu_{a}(x)=\Gamma(r)\mu_{0}\,,\label{TMux}
\end{align}
with a constant ratio of chemical potential to temperature throughout the system, given by $\alpha_{a}=\frac{\mu_{0}}{T_{0}}$.
Here, $T_{0}=\beta_{0}^{-1}$ and $\mu_{0}$ denote the temperature and chemical potential on the axis of rotation $r=0$, while both diverge at radius $r=\Omega^{-1}$ due to the Lorentz factor $\Gamma(r)=\frac{1}{\sqrt{1-\Omega^{2}r^{2}}}$.

Recalling \rf{GenP} in the context of pure rotation, we can express it in the following form:
\begin{align}
    &\int\,d^{3}\mathbf{x}\,\langle\HT^{k}_{k}(x)\rangle=-T_{0}\partial_{\lambda}\text{log}Z(\lambda)\Big|_{\lambda\rightarrow0}\,.\label{RotT}
\end{align}
In contrast to the case of homogeneous equilibrium, $\langle\HT^{k}_{k}(x)\rangle$ cannot be taken outside the integral due to its explicit coordinate dependence \cite{Ambrus:2019ayb,Ambrus:2019khr}. Owing to the inhomogeneous profile of the thermodynamic parameters as given in \rf{TMux}, the tensor densities become radially coordinate-dependent thermodynamic quantities. This can give rise to a scenario in which the volume-averaged tensor densities acquire an explicit dependence on the system size in the direction transverse to the rotation axis, denoted by $R$, the radius of the cylindrical rotating matter. 

In equilibrium, the condition of thermodynamic consistency entails extensivity, which follows from the assumption that thermodynamic variables (e.g., $O(x)$) vary over length scales much larger than the equilibrium correlation length. That is, 
$\lambda_{th}\partial O \ll L\partial O$, where $L$ and $\lambda_{th}$ correspond to the macroscopic scale ($\sim$ system size) and microscopic length scales (e.g., thermal de Broglie wavelength), respectively, thus establishing the scale hierarchy required for the applicability of hydrodynamic expansion~\cite{Kovtun:2022vas,Kovtun:2012rj}. For rigid rotation, the matter fields must be confined in the transverse direction relative to the rotation axis to ensure causality. Consequently, the radial extent $R$ must satisfy the condition $\Omega R\leq 1$. No such constraint applies to the extension $L_{z}$ of matter along the rotation axis, and it is assumed that the system is sufficiently extended in the longitudinal direction to meet the hydrodynamic hierarchy, i.e., $L_{z}\gg\lambda_{th}$. Similarly, along the radial direction, the hydrodynamic scale separation is maintained by imposing the condition $\lambda_{th}\ll R$ \cite{Bhattacharyya:2007vs,Kovtun:2022vas,Liu:2018kfw}-which excludes scenarios where the radial extent of the system is $\lesssim\lambda_{th}$- a situation that typically arises under conditions of very fast rotation and low temperatures. 

At this point, we focus on evaluating the left-hand side of \rf{RotT}, examining whether the volume-integrated space-like trace of the energy-momentum tensor can be expressed in terms of the extensive variable $L_{z}$ (or $V=\pi R^{2}L_{z}$) and $R$, along with the intensive thermodynamic parameters of the system. In this work, we consider a rotating gas of massless fermions. The computation is based on the statistically averaged energy-momentum tensor, using the expressions  derived in Refs.~\cite{Ambrus:2019ayb, Ambrus:2019khr}, where the results are provided in polar coordinates $(\mathbf{x},\mathbf{y},\mathbf{z})\rightarrow(r,\phi,\mathbf{z})$,
\begin{align}
    \langle\hat{T}^{rr}(x)\rangle&=\Gamma^{4}(r)\Big[P_{0}+\frac{\Omega^{2}}{24}B_{0}\Big(\frac{4}{3}\Gamma^{2}(r)-\frac{1}{3}\Big)\Big]\nonumber\\
    &=\langle\hat{T}^{\mathbf{z}\mathbf{z}}(x)\rangle\,,\label{Tzzrr}\\
    \langle\hat{T}^{\phi\phi}(x)\rangle&=r^{-2}\Gamma^{4}(r)\Big[P_{0}\Big(4\Gamma^{2}(r)-3\Big)+\frac{\Omega^{2}}{24}B_{0}\Big(8\Gamma^{4}(r)\nonumber\\
    &\quad\quad-8\Gamma^{2}(r)+1\Big)\Big]\,,\label{Tpp}\\
    \langle\hat{T}^{t\phi }(x)\rangle&=\Gamma^{6}(r)\Omega\Big[4P_{0}+B_{0}\frac{2\Omega^{2}}{9}\Big(\frac{3}{2}\Gamma^{2}(r)-\frac{1}{2}\Big)\Big]\,.\label{Tpht}\\
    \langle\hat{T}^{tt}(x)\rangle&=\Gamma^{4}(r)\Big[P_{0}\Big(4\Gamma^{2}(r)-1\Big)+\frac{\Omega^{2}}{8}B_{0}\Big(\frac{8}{3}\Gamma^{4}(r)\nonumber\\
    &\quad\quad-\frac{16}{9}\Gamma^{2}(r)+\frac{1}{9}\Big)\Big]\,.\label{Ttt}\\
\end{align}
Here, the divergent vacuum expectation value has been subtracted and the renormalized expressions for the averaged energy-momentum tensor components are presented in \rf{Tzzrr} and \rf{Tpp}. In the limit $\Omega\rightarrow0$, one recovers the known result for the non-rotating, massless fermion case. The expression $P_{0}(T_{0},\alpha_{a})=gT^{4}_{0}\Big(\frac{7\pi^{2}}{360}+$ $\frac{\alpha_{a}^{2}}{12}+\frac{\alpha^{4}_{a}}{24\pi^{2}}\Big)$
can be identified as the isotropic pressure in the non-rotating limit, where $g$ denotes the degeneracy factor of the fermion particles. Additionally, we observe that the thermodynamic function $B_{0}(T_{0},\alpha_{a})=gT_{0}^{2}\Big(1+\frac{3}{\pi^{2}}\alpha_{a}^{2}\Big)$ has a magnitude comparable to $\sim \beta_{0}^{2}P_{0}$. Substituting the expressions \rf{Tzzrr} and \rf{Tpp} into the left-hand side of \rf{RotT}, we arrive at
\begin{align}
    \int d^{3}\mxb\,\langle\hat{T}^{k}_{k}(x)\rangle&=-\int d^{3}\mxb\Big(\langle\hat{T}^{rr}(x)\rangle+\langle\hat{T}^{\mathbf{z}\mathbf{z}}(x)\rangle\nonumber\\
    &\quad+r^{2}\langle\hat{T}^{\phi\phi}(x)\rangle\Big),\nonumber\\
    &=-3V P_{0}(T_{0},\alpha_{a}) \frac{\big(1-\frac{\Omega^{2}R^{2}}{3}\big)}{\big(1-\Omega^{2}R^{2}\big)^{2}}\nonumber\\
    &-\frac{V \Omega^2 B_0}{216 (1-\Omega^2 R^2)^3} \Big( 3 R^4 \Omega^4 - 6 R^2 \Omega^2 + 27 \Big)\label{TotP}
\end{align}
Compared to the non-rotating case, the final result acquires a dependence on the angular velocity $\Omega$ through the appearance of the common dimensionless combination $\Omega R$. Moreover, within the hydrodynamic regime, the magnitudes of the first and second terms in the previous expression differ significantly.
This is because, in massless theories (or, strongly coupled system), the thermal de-Broglie wavelength is $\lambda_{th}\sim\frac{1}{T(r)}\Gamma(r)=\frac{1}{T_{0}}$ \footnote{A parallel understanding of this condition follows from quasiparticle kinematics, by requiring that the distance $\Delta$ between two successive collisions in the rotating fluid be significantly smaller than the system size, i.e., $\Delta \ll R$. The connection between $\Delta$ and the mean free path $\lambda_{f}(T(r),\alpha_{a})$ is established via a Lorentz transformation from the static lab frame to the fluid rest frame, giving
\begin{equation}
    \Delta =\lambda_f(T(r), \alpha_a)\, \Gamma(r)\left(1 + \frac{v}{v_p^{\parallel}}\right)\,,\nonumber
\end{equation}
where $v$ is the fluid velocity and $v_p^{\parallel}$ is the component of the quasiparticle velocity parallel to $v$. Since $\lambda_f(T(r), \alpha_a)$ depends on the local temperature $T(r)$, conformal invariance implies $\lambda_f(T(r), \alpha_a) =\lambda_f(T_0, \alpha_a)/\Gamma(r)$, so that
\begin{equation}
    \Delta = \lambda_f(T_0, \alpha_a)\left(1 + \frac{v}{v_p^{\parallel}}\right)\,.\nonumber
\end{equation}
As the factor $\left(1 + v/v_p^{\parallel}\right)$ lies between $1$ and $2$, the scale hierarchy is equivalently expressed as $\Delta\sim\lambda_{f}(T_{0},\alpha_{a})\ll R$. More generally, the mean free path $\lambda_f$ can be replaced by a microscopic correlation length, such as the thermal de Broglie wavelength $\lambda_{\text{th}}$.}, where $\Gamma(r)$ appears as a consequence of the Lorentz transformation from the static coordinate frame and the local fluid rest frame \cite{Dey:2024crk,Bhattacharyya:2007vs}.
Therefore, to ensure the validity of a hydrodynamic description, a stringent scale hierarchy must be imposed by the condition $ R\gg\frac{1}{T_{0}}$ (or equivalently, $\Omega R\gg\frac{\Omega}{T_{0}}$ ). This implies that the terms involved with $\Omega R$ dominate over those involved with $\beta_{0}\Omega$ in thermodynamic quantities. As a result, the second term in Eqs.~\eqref{Tzzrr}-\eqref{TotP} can be neglected compared to the first term because $\Omega^{2}B_{0}\sim(\Omega\beta_{0})^{2}P_{0}\ll P_{0}$.

Revisiting the left-hand side of \rf{RotT}, we observe that the derivative with respect to the scaling parameter $\lambda$ receives contributions from each extensive variable present in the argument of $\text{log}Z$, provided that they attribute to the scaling of the thermodynamic potential, that is,
\begin{align}
    \partial_{\lambda}\text{log}Z(\lambda)\big|_{\lambda\rightarrow0}&=L_{z}\partial_{L_{z}}(\log Z)\big|_{R}+R\partial_{R}(\log Z)\big|_{L_{z}}\nonumber\\
    &=3V\partial_{V}(\text{log}Z)\big|_{\Omega R}+\Omega R\partial_{\Omega R}(\text{log}Z)\big|_{V}\,.\label{delVZ}
\end{align}
It is important to note that the dependence $R$ appears in the expression on the right hand side of Eq.~\eqref{TotP}\,
 through the dimensionless combination $\Omega R$. Therefore, an ansatz for $\text{log}Z$ can be proposed, incorporating the extensivity property
\begin{align}
    \text{log}Z=\beta_{0}\mathcal{P}(T_{0},\alpha_{a},\Omega R)V\,.\label{Ansatz}
\end{align}
Here, the minus sign is introduced by convention.
However, the precise functional form of $\mathcal{P}$ depends on the specific thermodynamic system under consideration. Nevertheless, the extensivity of $\text{log}Z$ is due to the scaling of volume $V$ as well as through dependency  of transverse size $R$.
This is further supported by the expression in \rf{TotP}, where the dependence on $R$ (as well as on $\Omega$) appears exclusively through the dimensionless common product $\Omega R$.
Therefore, the average pressure $\mathcal{P}(T_{0},\alpha_{a},\Omega,R)$ can be better expressed as $\mathcal{P}(T_{0},\alpha_{a},\Omega R)$, and differentiation over $\text{log}Z(\lambda)$ in the left-hand side of the \rf{RotT} can be written as
\begin{align}
\partial_{\lambda}\text{log}Z(\lambda)\big|_{\lambda\rightarrow0}&=\beta_{0}V\big(3+\Omega R\,\partial_{\Omega R}\big)\mathcal{P}(\beta_{0},\alpha_{a},\Omega R) \label{delVZ1}
\end{align}
Substituting the above expression and Eq.~\eqref{TotP} into Eq.~\eqref{RotT}, one can derive the expression for $\mathcal{P}(T_{0},\alpha_{a},\Omega R)$ in the following form:
\begin{align}
    \mathcal{P}(T_{0},\alpha_{a},\Omega R) =  \frac{P_0(T_0,\alpha_a)}{(1-\Omega^2 R^2)}
+\frac{B_0 \Omega^2 (3 - R^2 \Omega^2)}{72 (1-R^2\Omega^2)^2},\label{Ptot}
\end{align}
which, upon using Eq.~\eqref{Thpot}, gives the thermodynamic potential in the form
\footnote{ The dependency on the system size in the effective average pressure can occur with non-uniform pressure which has a non-zero gradient to oppose the external force, like for an isothermal atmosphere the pressure decreases exponentially with altitude $z$ in presence of constant gravitational acceleration $g$, expressed as $P(z)=P_{0}e^{-mg z\beta}$, where $\beta$, and $m$ and $g$ are the inverse of temperature, the mass of the molecules of each strata of the atmosphere. Where averaged pressure $\mathcal{P}=\frac{P_{0}}{mgh\beta}(1-e^{-mg h\beta})$ for an air column with vertical extent $h$. Here, the vertical extent appears through the combination $ m g h\beta$ in volume-averaged pressure $\mathcal{P}$. }:
\begin{align}
    \Phi(T_{0},\alpha_{a},\Omega R) =  -V\Bigg(\frac{P_0(T_0,\alpha_a)}{(1-\Omega^2 R^2)}
+\frac{B_0 \Omega^2 (3 - R^2 \Omega^2)}{72 (1-R^2\Omega^2)^2}\Bigg)\,.
\end{align}
Using the hierarchy of hydrodynamic scales $R\gg\beta_{0}$, one obtains, at leading order, an expression that matches Ref.~\cite{Bhattacharyya:2007vs}, namely,
\begin{align}
    \Phi(T_{0},\alpha_{a},\Omega R,V)
    &=-V\frac{P_{0}(T_{0},\alpha_{a})}{\big(1-\Omega^{2}R^{2}\big)}+\mathcal{O}(\Omega^{2}\beta_{0}^{2})\,,\label{PhiRot}
\end{align}
and, similarly, the corresponding leading-order contribution to the pressure can be expressed as $\mathcal{P}(T_{0},\alpha_{a},\Omega R)=\mathcal{P}^{0}(T_{0},\alpha_{a},\Omega R)+\mathcal{O}(\Omega^{2}\beta_{0}^{2})$, where the leading-order pressure is given by
\begin{align}
   \mathcal{P}^{0}(T_{0},\alpha_{a},\Omega R) =\frac{P_{0}(T_{0},\alpha_{a})}{\big(1-\Omega^{2}R^{2}\big)}\,,\label{PRot}
\end{align}
which also captures the pressure anisotropy induced by rotation, as follows:
\begin{align}
    \mathcal{P}_{\perp}&=-\frac{R}{2V}\partial_{R}\Phi=\frac{\mathcal{P}^{0}(T_{0},\alpha_{a},\Omega R)}{(1-\Omega^2 R^2)}+\mathcal{O}(\Omega^{2}\beta_{0}^{2})\,\,\,,\nonumber\\
    \mathcal{P}_{||}&=-\frac{L_{z}}{V}\partial_{L_{z}}\Phi=\mathcal{P}^{0}(T_{0},\alpha_{a},\Omega R)+\mathcal{O}(\Omega^{2}\beta_{0}^{2})\,.
\end{align}
Here, $\mathcal{P}_{\perp}$ and $\mathcal{P}_{||}$ are the pressure along the perpendicular and parallel to rotation axis, respectively. So, this helps us to understand that the volume averaged hydrodynamic pressure as expressed in the left hand side of Eq.~\eqref{RotT} can be written as $\mathcal{P}_{||}+2\mathcal{P}_{\perp}$.
Following the Eq.~\eqref{PhiRot}, one observes that the dependence on $\Omega$ (or, R) enters the thermodynamic potential $\Phi$ through the common product $\Omega R$. This structure carries over to all thermodynamic quantities, as required by thermodynamic consistency. For example, the thermodynamic quantities such as the entropy $S$, total angular momentum $\vec{J}$ along the rotation axis $\hat{n}=\frac{\vec{\Omega}}{\Omega}$, and the conserved charges $Q_{a}$ can be extracted from the thermodynamic potential (See Appendix~\ref{sec:comGLW}). For the present, the total angular momentum carried by this rigidly rotating matter has been calculated \footnote{ By using the expectation value of the $(t,\phi)$ component of the energy-momentum tensor as expressed in Eq.~\eqref{Tpht}, one can calculate the angular momentum as $\vec{J}=\int d^{3}\mathbf{x}\langle\hat{T}^{t\phi}(x)\rangle$, which is consistent with the expression Eq.~\eqref{JRot} in the hydrodynamic regime.  }
 \begin{align}
     \langle\hat{J}^{xy}\rangle&=-T_{0}\partial_{\Omega}\text{log}Z\Big|_{T_{0},\alpha_{a}}\nonumber\\
     &=V\frac{2\Omega R^{2}P_{0}(T_{0},\alpha_{a})}{(1-\Omega^{2}R^{2})^{2}}+V\frac{B_0\Omega (3+R^2\Omega^2)}{36(1-R^2\Omega^2)^3}\,,\label{JRot}
 \end{align}
which allows us to compute the isothermal moment of inertia along the principal axis of rotation, taken to be $\hat{n}=\frac{\vec{\Omega}}{\Omega}$ (assumed, without loss of generality, to lie along the $z$-axis), given by
\begin{align}
    I(\Omega)&=\frac{\langle\hat{J}^{xy}\rangle}{\Omega}\,,\nonumber\\
    &=V\Bigg(\frac{2 R^{2}P_{0}(T_{0},\alpha_{a})}{(1-\Omega^{2}R^{2})^{2}}+\frac{B_0 (3+R^2\Omega^2)}{36(1-R^2\Omega^2)^3}\Bigg)\label{MI}
\end{align}
Using the hierarchy of hydrodynamic scales $R\gg\beta_{0}$, we obtain the leading-order contribution:
\begin{equation}
I(\Omega) = \frac{2 P_0 R^2}{(1-R^2\Omega^2)^2} + \mathcal{O}(\beta^{2}\Omega^{2})\,.\label{MI1}
\end{equation}
Moreover, we can also define the isothermal differential moment of inertia, $\tilde{I}=\frac{\partial \langle\hat{J}^{xy}\rangle}{\partial\Omega}$,whose positivity ensures the stability of the rotating system. Using Eq.~\eqref{JRot}, one can show that its leading-order contribution is given by
\begin{align}
\tilde{I}(\Omega)&=\partial_{\Omega}\langle\hat{J}^{xy}\rangle\nonumber\\
    &=V\frac{2R^{2}P_{0}(T_{0},\alpha_{a})}{(1-\Omega^{2}R^{2})^{3}}\Big(1+3\Omega^{2}R^{2}\Big)+\mathcal{O}(\Omega^{2}\beta_{0}^{2})\,,\label{IRot}
\end{align}
which also helps identify how the difference between the pressures along the radial direction and parallel to the rotation axis is related to the rotational effect, namely $\mathcal{P}_{\perp}-\mathcal{P}_{||}=\tilde{I}\frac{\Omega^{2}}{2V}+\mathcal{O}(\beta^{2}_{0}\Omega^{2})$.

Furthermore, in the static limit $\lim_{\Omega\rightarrow0}$, both types of moment of inertia defined in Eqs.~\eqref{MI} and \eqref{IRot} converge to the same expression, i.e., $\lim_{\Omega\rightarrow0} I(\Omega)=\lim_{\Omega\rightarrow0}\tilde{I}(\Omega)=I_{s}$, which is expressed as
\begin{align}
I_{s}=2VP_{0}(T_{0},\alpha_{a})R^{2}+\frac{V B_0}{12}\,.\label{IsRot}
\end{align}
This result is also consistent with previous studies \cite{Patuleanu:2025zbn, Bhattacharyya:2007vs}. One can identify that, at the hydrodynamic scale, the term proportional to $B_{0}$ arises from a quantum correction due to the spin contribution to the moment of inertia \cite{Patuleanu:2025zbn}, which is suppressed.
Here, both isothermal moments of inertia remain positive within the range $\Omega R\leq1$, ensuring rotational stability for the thermal system \cite{weinhold1975metric,weinhold1975metric,landau1984statistical}. It is also worth noting that the moment of inertia $I(\Omega)$ at non-vanishing $\Omega$ is larger than its static value $I_{s}$, as rotation tends to redistribute matter radially outward. In comparison to Refs.~\cite{Braguta:2023tqz,Braguta:2023yjn,Braguta:2023kwl,PhysRevD.107.114502,Braguta:2020biu}, one can expand the thermodynamic potential under the assumption of slow rotation, i.e., $\Omega R\ll1$, yielding 
\begin{align}
    \Phi(T_{0},\alpha_{a},\Omega R, V)&=-V P_{0}\Big(1+\Omega^{2}R^{2}\Big)+\mathcal{O}(\Omega^{4}R^{4})\,,\nonumber\\
    &=\Phi(T_{0},\alpha_{a},V)-I_{s}\frac{\Omega^{2}}{2}\,.
\end{align}
From the last expression, one can identify the moment of inertia as $I_{s}$, which remains independent of the angular velocity $\Omega$, which implies the system undergoes no shape deformation.

On a final note, this work has demonstrated how the principle of extensivity governs the structure of the thermodynamic potential in homogeneous and inhomogeneous settings, thereby ensuring consistency with the thermodynamic limit. In the presence of rigid rotation, the thermodynamic pressure $\mathcal{P}(\beta_0, \alpha_a, \Omega R)$ depends on the angular velocity $\Omega$ only through the dimensionless combination $\Omega R$, and remains finite as long as the system size satisfies $R < \Omega^{-1}$.

It is worth noting that, in general, the $\Omega$-dependence of the pressure may enter through either $\Omega/T_0$ or $\Omega R$, as seen from the right-hand side of Eq.~\eqref{TotP}. However, within the hydrodynamic regime, the hierarchy $\Omega R \gg \Omega/T_0$ must be maintained, thus justifying the retention of only the contributions involving $\Omega R$ in thermodynamic quantities.

\section{Conclusions}
\label{conclusion}

Here, an alternative real-time method has been described, based on the ETC relations between the dilatation generator and both the global charges and their associated currents at global equilibrium, as discussed in Sec.~\ref{SecVirial}. It is also known that the definitions of Poincaré charges given in \rf{PVB} and \rf{JVB} are independent of the particular choice of the energy-momentum tensor, provided the different forms are related by the pseudo-gauge transformations defined in Eqs.~\eqref{PGTT} and \eqref{PGST}. In addition, note that, apart from the Schwinger terms in the ETCs, the remaining terms contain a single derivative on the Dirac delta functions. These terms are invariant under pseudo-gauge transformations, and these structures are not affected by the particular choice of the matter field with spin $\leq 1$~\cite{Schwinger:1963zz,Landsman:1986uw,Deser:1967zzf,Trubatch:1970zr,Jackiw:1968lmf}. From the general structure of equal-time commutators (ETCs), we derive Eq.~\eqref{GenP}, which captures the scaling behavior of $\log(Z)$ in theories of renormalizable matter fields with spin $\leq 1$, excluding gauge fields.
The exclusion of the gauge field occurs due to the particular choice of the dilatation generator $\hat{D}(t)$ through \rf{DiLI} instead of \rf{DilC}, which involves the improved energy-momentum tensor $\hat{T}^{\mu\nu}$ to investigate the scaling properties (here it is a rigid scaling transformation) of the systems considered~\cite{francesco2012conformal,Callan:1970ze,Coleman:1970je,PhysRevD.2.1541}. In our calculation, we have considered fermionic particles of spin $1/2$. Thus, following the definition of the dilatation generator, \rf{DiLI} has been used to compute thermodynamic variables, e.g. moment of inertia. 

With homogeneous thermodynamic parameters, it has been shown that the thermodynamic potential $\Phi$ does not have any dependence on the size of the system other than volume term, $V$. In the case of rigid rotation, the dependency on the radial extent $R$ appears through the common product $\Omega R$ (see \rf{PhiRot}). In recent results from rotating $SU(3)$ lattice gauge theory \cite{Braguta:2023tqz,Braguta:2023yjn,Braguta:2023kwl,PhysRevD.107.114502,Braguta:2020biu} as well as some thermal field theory calculations \cite{PhysRevD.103.094515,Chernodub:2016kxh, Ambrus:2023bid}, the dependence of the transverse size is pronounced through the presence of a common product $\Omega R$. However, in some theoretical approaches, the dependence on angular frequency $\Omega$ arises from a different common product $\frac{\Omega}{T_{0}}$ \cite{Huang:2020dtn,Fujimoto:2021xix}. Here, it has been shown that the condition for the description of the applicability of hydrodynamics can be ensured by the hierarchy $R\gg\frac{1}{T_{0}}$, which enforces the dominance of $\Omega R $ terms in thermodynamic quantities. Consequently, the term containing $B_{0}\Omega^{2}$ ($\sim P_{0}\beta_{0}^{2}\Omega^{2}$) in Eqs.~\eqref{Tzzrr} and \eqref{Tpp}, can be neglected compared to the magnitude of $P_{0}$. This has been taken care of, to express the intensive pressure $\mathcal{P}$ as given in \rf{PhiRot} from \rf{Ptot}, along with the extensivity properties of $\text{log}Z$.

The \rf{MI1} of this study can be useful in understanding the contribution of quarks to the moment of inertia of QGP produced in heavy-ion collisions, which has the greatest vorticity of any fluid produced in a laboratory \cite{STAR:2017ckg,Becattini:2020ngo,Huang:2020dtn,Wang:2017jpl}.  
As an example, for a vortical QGP medium, one may consider the following values of the physical parameters: $T_0 \sim 200$MeV, $\Omega \sim10^{22}s^{-1}\approx7$MeV. In particular, for Au+Au collisions, one may also consider $\mu_0 \sim 0.5~\mathrm{fm}^{-1}c$, with a system size $R \sim 14~\mathrm{fm}$~\cite{Becattini:2020ngo}.
Using this set of parameters, the condition for the validity of hydrodynamics in such systems is found to be valid as the hierarchy $\Omega^{2}R^{2}\sim 300\Omega^{2}\beta_{0}^{2}\gg\Omega^{2}\beta_{0}^{2}$ is maintained. Hence, the dependency on the angular frequency $\Omega$ in the thermodynamic variables from the contribution of $\Omega \beta_{0}$ is suppressed. Moreover, with this consideration, the moment of inertia has been computed for rigidly rotating massless fermionic matter, as given in \rf{MI1}. The qualitative estimation of the light quarks contribution to the QGP, $I\sim 4.15 I_{s}$ at $T_{0}\geq T_{c}$ and $\Omega R\sim 0.5$, where the values of the static limit of moment of inertia are $I_{s}\approx 1.08\times10^{3}\text{MeV}^{-1}$ and $I_{s}\approx 1.62\times10^{3} \text{MeV}^{-1}$ for $N_{f}=2$ (or $g=4$) and $N_{f}=3$ (or $g=6$), respectively. The computed moment of inertia agrees with previously reported results \cite{Patuleanu:2025zbn, Bhattacharyya:2007vs}. However, it differs from the value obtained in \cite{Ahadi:2025rqs}, where the moment of inertia is associated with the axial current density and is pseudo-gauge dependent. In global equilibrium, the result should not depend on the pseudo-gauge, since the density matrix is pseudo-gauge invariant as it is constructed from total conserved charges as shown in Eq.~\eqref{rhodenT0Om}. In our approach, the moment of inertia is unaffected by ambiguities related to the separation of spin and orbital components because it is derived directly from the thermodynamic potential. Moreover, the moment of inertia associated with a local current is more appropriately interpreted as the vortical conductivity of the axial current, as discussed in \cite{Patuleanu:2025zbn}. 

Along these lines, an interesting extension would be to establish an analytical estimation of the moment of inertia for gluon fields using the real-time method described in Sec.~\ref{SecVirial}.
An investigation has recently been conducted from the first-principle lattice calculation \cite{Braguta:2023yjn,Braguta:2023tqz} with the help of imaginary time formalism, and an interesting result was derived for QGP having a negative moment of inertia below a particular temperature $\sim1.5 T_{c}$. However, the approach in Sec.~\ref{SecVirial} excludes the gauge field. It can be generalised for the gauge field by introducing the appropriate virial field in \rf{DilC}. For example, $SU(N)$ Lagrangian for the non-abelian gauge field $A^{\mu}_{a}$ yields $V^{\mu}=2\eta A^{\mu}_{a}\partial\cdot A_{a}-\partial^{\mu}(\bar{\xi}_{a}\xi_{a})-g f_{abc}\bar{\xi}_{a}A^{\mu}_{b}\xi_{c}$, where $\eta$, $g$, and $\xi_{a}$ are the gauge parameter, coupling constant, and ghost field, respectively. The presence of this virial field in the dilatation generator $\hat{D}(t)$ in Eq.~\eqref{DilC} modifies the ETC given in \rf{DT00C} and \rf{DT0iC} only by the BRS-transformed operator\cite{Landsman:1986uw,Freedman:1974gs,Pascual:1984zb}. However, this extra BRS-transformed part of an operator vanishes on the grand-canonical averages, which involves the projection of the physical states \cite{Hata:1980yr,Ojima:1981ma}. So, in the context of extracting thermodynamic quantities for gauge theory, the 
``improved" tensor $\hat{T}^{\mu\nu}$ can be replaced by the Belinfante tensor $\hat{T}_{B}^{\mu\nu}$ in \rf{GenP} because it is indistinguishable from the physical thermal states. A separate study with the gauge field will be reported elsewhere.

The present approach relies on the extensivity of $\log Z$, although this scaling may be altered by quantum corrections. In particular, conformal (trace) anomalies \cite{Adler:1976zt, Collins:1976yq} are expected to modify the formalism, and exploring this extension constitutes an interesting avenue for future investigation.

Moreover, in the context of (non-)relativistic many-body systems, spin hydrodynamics has been kept alive due to several studies \cite{PhysRevB.96.020401,PhysRevB.104.184414,takahashi2016spin,PhysRevLett.106.076601,Bhadury:2020puc,Bhadury:2022ulr,Weickgenannt:2024esg,Sheng:2021kfc}. In particular, recent studies of the spin polarization of hadrons produced from QGP \cite{Becattini:2022zvf,Becattini:2021lfq}, and in condensed matter, the spin voltage arises due to the vorticity of the flowing spin liquids \cite{takahashi2016spin} drawing attention to theoretical investigations~(see Refs.~\cite{Voloshin:2004ha,Becattini:2007nd,Becattini:2009wh, Montenegro:2017rbu, Li:2017slc, Ayala:2019iin,Wang:2017jpl,Florkowski:2017ruc,Bhadury:2022ulr, Florkowski:2017dyn,Florkowski:2018fap,Weickgenannt:2019dks, Kapusta:2019sad,Weickgenannt:2020aaf,Speranza:2020ilk,Shi:2020htn,Bhadury:2020cop,Hu:2021lnx,Bhadury:2021oat,Fu:2021pok,Becattini:2021suc, Becattini:2021iol,Hongo:2021ona,Weickgenannt:2022zxs,Gallegos:2022jow,Bhadury:2022ulr,Weickgenannt:2022jes,Weickgenannt:2022qvh,Sarwar:2022yzs,Biswas:2022bht,Biswas:2023qsw,Banerjee:2024xnd,Wagner:2024fhf,Lin:2024cxo,Bhadury:2024ckc,Bhadury:2025vvo,Bhadury:2025fil,Buzzegoli:2024mra,Sapna:2025yss,Weickgenannt:2024ibf,Dey:2023hft,Das:2025kgq}).
In addition, the emergence of the hydrodynamic regime in graphene \cite{Lucas:2017idv} could be helpful to understand the magnetovortical effect by establishing a tabletop experiment with a proper initial inhomogeneous temperature profile as given in \rf{TMux}, and allows for the conversion of the orbital angular momentum to the spin angular momentum via the spin orbital effect, as derived in \cite{Jaiswal:2024urq}. In this article, the study of the rotating system in global equilibrium shows that the total angular momentum can be computed as given in \rf{JRot} along the principal axis of the system. However, if shape deformation leads to a change (slowly) of the principal axis that gives the spin polarization in the presence of the spin-orbit coupling, the system becomes magnetized. Also, the rigid rotation of the matter discards the effect of shear and bulk viscosity, but with the non-equilibrium statistical method, one can take into account the off-equilibrium contribution in the density matrix $\hat{\rho}$ with hydrodynamic gradient expansions, which may lead to some effect on spin transport. Both of these aspects will be explored in the forthcoming studies.

\section{Acknowledgment}
\label{sec:ack}

S.D. acknowledges Hiranmaya Mishra, Amaresh Jaiswal, Arpan Das, Samapan Bhadury, and Victor E. Ambrus for valuable discussions, encouragement, and critical reading of the manuscript.

\appendix
\section{Thermodynamic Relations}
\label{sec:comGLW}
In this appendix, thermodynamics-consistent relations have been derived using the density operators \rf{rhoden}, starting with the definition of entropy $S$ as
\begin{align}
    S&=-\langle\text{log}\hat{\rho}\rangle\,,\nonumber\\
    &=b_{\mu}\langle\hat{P}^{\mu}\rangle-\frac{1}{2}\omega_{\mu\nu}\langle\hat{J}^{\mu\nu}\rangle-\alpha_{a}\langle\hat{Q}_{a}\rangle+\text{log}Z\,.\label{En}
\end{align}
Here, $b_{\mu},\omega_{\mu\nu}$ and $\alpha_{a}$ are the Lagrange multipliers for the corresponding global charges $\hat{P}^{\mu},\hat{J}^{\mu\nu}$ and $\hat{Q}_{a}$, respectively. Where $Z$ is the partition function defined in \rf{PartN}.
Thus one can obtain 
\begin{align}
    \delta S=b_{\mu}\langle\delta\hat{P}^{\mu}\rangle-\frac{1}{2}\omega_{\mu\nu}\langle\delta\hat{ J}^{\mu\nu}\rangle-\alpha_{a}\langle\delta\hat{Q}_{a}\rangle\,,\label{fLawTher}
\end{align}
which is the first law of thermodynamics.

Turning to solutions of the global equilibrium condition \rf{KillCon}, one can identify as $b_{0}=\frac{1}{T_{0}}=\beta_{0},b_{i}=\frac{\mathbf{V}}{T_{0}}$. Where, $\omega_{\mu\nu}$ becomes $\frac{\mathbf{a}}{T_{0}}(\delta^{\mu}_{0}\delta^{\nu}_{\mathbf{z}}-\delta^{\mu}_{\mathbf{z}}\delta^{\nu}_{0})$ and $\frac{\Omega}{T_{0}}(\delta^{\mu}_{\mxb}\delta^{\nu}_{\mathbf{y}}-\delta^{\mu}_{\mathbf{y}}\delta^{\nu}_{\mxb})$ if the fluid is accelerated and rotating along the $\mathbf{z}$ axis, where $T_{0},\mathbf{V},\mathbf{a}$ and $\Omega$ are temperature, center-of-mass velocity, acceleration, and angular velocity, respectively. However, $\mu_{a}=T_{0}\alpha_{a}$ is the chemical potential for each corresponding conserved matter charge.

More specifically, if we rewrite the \rf{En} for the rigid rotation solution with the killing vector 
\begin{align}
   b_{\mu}=(\beta_{0},\mathbf{0})\,,\text{and}\, \omega_{\mu\nu}=\beta_{0}\Omega(g_{\mu x}g_{\nu y}-g_{\mu y}g_{\nu x})\,,
\end{align}
which helps to rewrite the \rf{En} in this context as
\begin{align}
    T_{0}S=E-\vec{\Omega}\cdot\vec{J}-\mu_{a}Q_{a}+T_{0}\text{log}Z\,.
\end{align}
Here, $E=\langle P^{0}\rangle$, $\vec{J}=\langle \hat{J}\rangle$, and $Q_{a}=\langle\hat{Q}_{a}\rangle$ are the energy, angular momentum along the principal axis, and the values of charges, respectively, computed from the average of the grand canonical ensemble at global equilibrium with the density operator $\hat{\rho}$.
Define
\begin{align}
    \Phi&=-T_{0}\text{log}Z\,.\label{Thpot}
\end{align}
The last equality of the last expression is the consequence of the exetensivity property of the thermodynamic potential $\Phi$ under the scaling $\mxb\rightarrow e^{\lambda}\mxb$ as $\Phi\rightarrow e^{3\lambda}\Phi$, thus $\mathcal{P}$ is an intensive variable (See \rf{Ansatz}) \cite{Vilenkin:1980zv,Bhattacharyya:2007vs}. Applying the first law of thermodynamics \rf{fLawTher}, one may arrive at
\begin{align}
    \delta \Phi=-\delta T_{0} S-\delta\vec{\Omega}\cdot\vec{J}-\delta\mu_{a} Q_{a}\,,
\end{align}
which defines the thermodynamics variables entropy $S$, angular momentum $\vec{J}$, and charges $Q_{a}$ as
\begin{align}
    S&=-\bigg(\frac{\partial\Phi}{\partial T_{0}}\bigg)_{\vec{\Omega},\mu_{a}}\,,\label{entropy}\\
    \vec{J}&=-\bigg(\frac{\partial\Phi}{\partial \vec{\Omega}}\bigg)_{T_{0},\mu_{a}}\,,\label{angularmomentum}
\end{align}
and
\begin{align}
    Q_{a}=-\bigg(\frac{\partial\Phi}{\partial \mu_{a}}\bigg)_{T_{0},\vec{\Omega}}\,.\label{charge}
\end{align}
Using these definitions for a given thermodynamic potential $\Phi$ \rf{PhiRot} can be shown as 
\begin{align}
    S=V\frac{\partial P_{0}}{\partial T_{0}}\frac{\big(1-\frac{\Omega^{2}R^{2}}{3}\big)}{\big(1-\Omega^{2}R^{2}\big)^{2}}+\mathcal{O}(\beta^{2}\Omega^{2})\,,
\end{align}
\begin{align}
    Q_{a}=V\frac{\partial P_{0}}{\partial \mu_{a}}
    \frac{\big(1-\frac{\Omega^{2}R^{2}}{3}\big)}{\big(1-\Omega^{2}R^{2}\big)^{2}}+\mathcal{O}(\beta^{2}\Omega^{2})\,,
\end{align}
and the angular momentum along the principle axis $\vec{J}$ has been calculated in the \rf{JRot}.

\bibliography{ref}
\end{document}

%% file: Command.tex
\def\be{\begin{equation}}
	\def\ee{\end{equation}}
\newcommand{\bel}[1]{\begin{eqnarray}\label{#1}}
	\newcommand{\eel}{\end{eqnarray}}
\def\barr{\begin{array}}
	\def\earr{\end{array}}
\def\beq{\begin{eqnarray}}
	\def\eeq{\end{eqnarray}}
\def\bfig{\begin{figure}}
	\def\efig{\end{figure}}

\newcommand{\rf}[1]{Eq.~(\ref{#1})}

\newcommand{\bea}{\begin{eqnarray}}
\newcommand{\eea}{\end{eqnarray}}

\newcommand{\av}{{\boldsymbol a}} 
\newcommand{\bv}{{\boldsymbol b}} 

\def\S0iU{{\Sigma}^{0i}}

\def\n0{n_{(0)}}
\def\e0{\varepsilon_{(0)}}
\def\P0{P_{(0)}}

\def\PV{\hat{P}}
\def\JV{\hat{J}}
\def\HT{\hat{T}}
\def\HJ{\hat{J}}
\def\HS{\hat{S}}
\def\HD{\hat{D}}
\def\mxb{\mathbf{x}}

%% file: ref.bib
@article{Weickgenannt:2022jes,
    author = "Weickgenannt, Nora and Wagner, David and Speranza, Enrico",
    title = "{Pseudogauges and relativistic spin hydrodynamics for interacting Dirac and Proca fields}",
    eprint = "2204.01797",
    archivePrefix = "arXiv",
    primaryClass = "nucl-th",
    doi = "10.1103/PhysRevD.105.116026",
    journal = "Phys. Rev. D",
    volume = "105",
    number = "11",
    pages = "116026",
    year = "2022"
}

@article{c,
  title = {Pseudogauges and relativistic spin hydrodynamics for interacting Dirac and Proca fields},
  author = {Weickgenannt, Nora and Wagner, David and Speranza, Enrico},
  journal = {Phys. Rev. D},
  volume = {105},
  issue = {11},
  pages = {116026},
  numpages = {18},
  year = {2022},
  month = {Jun},
  publisher = {American Physical Society},
  doi = {10.1103/PhysRevD.105.116026},
  url = {https://link.aps.org/doi/10.1103/PhysRevD.105.116026}
}

@article{Becattini:2012pp,
    author = "Becattini, F. and Tinti, L.",
    title = "{Nonequilibrium Thermodynamical Inequivalence of Quantum Stress-energy and Spin Tensors}",
    eprint = "1209.6212",
    archivePrefix = "arXiv",
    primaryClass = "hep-th",
    doi = "10.1103/PhysRevD.87.025029",
    journal = "Phys. Rev. D",
    volume = "87",
    number = "2",
    pages = "025029",
    year = "2013"
}

@article{Becattini:2011ev,
    author = "Becattini, F. and Tinti, L.",
    title = "{Thermodynamical inequivalence of quantum stress-energy and spin tensors}",
    eprint = "1101.5251",
    archivePrefix = "arXiv",
    primaryClass = "hep-th",
    doi = "10.1103/PhysRevD.84.025013",
    journal = "Phys. Rev. D",
    volume = "84",
    pages = "025013",
    year = "2011"
}

@article{Weickgenannt:2019dks,
    author = "Weickgenannt, Nora and Sheng, Xin-Li and Speranza, Enrico and Wang, Qun and Rischke, Dirk H.",
    title = "{Kinetic theory for massive spin-1/2 particles from the Wigner-function formalism}",
    eprint = "1902.06513",
    archivePrefix = "arXiv",
    primaryClass = "hep-ph",
    doi = "10.1103/PhysRevD.100.056018",
    journal = "Phys. Rev. D",
    volume = "100",
    number = "5",
    pages = "056018",
    year = "2019"
}

@article{Speranza:2020ilk,
    author = "Speranza, Enrico and Weickgenannt, Nora",
    title = "{Spin tensor and pseudo-gauges: from nuclear collisions to gravitational physics}",
    eprint = "2007.00138",
    archivePrefix = "arXiv",
    primaryClass = "nucl-th",
    doi = "10.1140/epja/s10050-021-00455-2",
    journal = "Eur. Phys. J. A",
    volume = "57",
    number = "5",
    pages = "155",
    year = "2021"
}

@article{Voloshin:2004ha,
      author         = "Voloshin, Sergei A.",
      title          = "{Polarized secondary particles in unpolarized high energy
                        hadron-hadron collisions?}",
      year           = "2004",
      eprint         = "nucl-th/0410089",
      archivePrefix  = "arXiv",
      primaryClass   = "nucl-th",
      SLACcitation   = "%%CITATION = NUCL-TH/0410089;%%"
}

@article{Becattini:2007nd,
      author         = "Becattini, F. and Piccinini, F.",
      title          = "{The Ideal relativistic spinning gas: Polarization and
                        spectra}",
      journal        = "Annals Phys.",
      volume         = "323",
      year           = "2008",
      pages          = "2452-2473",
      doi            = "10.1016/j.aop.2008.01.001",
      eprint         = "0710.5694",
      archivePrefix  = "arXiv",
      primaryClass   = "nucl-th",
      SLACcitation   = "%%CITATION = ARXIV:0710.5694;%%"
}

@article{Becattini:2007sr,
      author         = "Becattini, F. and Piccinini, F. and Rizzo, J.",
      title          = "{Angular momentum conservation in heavy ion collisions at
                        very high energy}",
      journal        = "Phys. Rev.",
      volume         = "C77",
      year           = "2008",
      pages          = "024906",
      doi            = "10.1103/PhysRevC.77.024906",
      eprint         = "0711.1253",
      archivePrefix  = "arXiv",
      primaryClass   = "nucl-th",
      SLACcitation   = "%%CITATION = ARXIV:0711.1253;%%"
}

@article{Li:2017slc,
      author         = "Li, Hui and Pang, Long-Gang and Wang, Qun and Xia,
                        Xiao-Liang",
      title          = "{Global $\Lambda$ polarization in heavy-ion collisions
                        from a transport model}",
      journal        = "Phys. Rev.",
      volume         = "C96",
      year           = "2017",
      number         = "5",
      pages          = "054908",
      doi            = "10.1103/PhysRevC.96.054908",
      eprint         = "1704.01507",
      archivePrefix  = "arXiv",
      primaryClass   = "nucl-th",
      SLACcitation   = "%%CITATION = ARXIV:1704.01507;%%"
}

@article{STAR:2017ckg,
      author         = "Adamczyk, L. and others",
      title          = "{Global $\Lambda$ hyperon polarization in nuclear
                        collisions: evidence for the most vortical fluid}",
      collaboration  = "STAR",
      journal        = "Nature",
      volume         = "548",
      year           = "2017",
      pages          = "62-65",
      doi            = "10.1038/nature23004",
      eprint         = "1701.06657",
      archivePrefix  = "arXiv",
      primaryClass   = "nucl-ex",
      SLACcitation   = "%%CITATION = ARXIV:1701.06657;%%"
}

@article{Betz:2007kg,
    author = "Betz, Barbara and Gyulassy, Miklos and Torrieri, Giorgio",
    title = "{Polarization probes of vorticity in heavy ion collisions}",
    eprint = "0708.0035",
    archivePrefix = "arXiv",
    primaryClass = "nucl-th",
    doi = "10.1103/PhysRevC.76.044901",
    journal = "Phys. Rev. C",
    volume = "76",
    pages = "044901",
    year = "2007"
}

@article{Becattini:2009wh,
      author         = "Becattini, F. and Tinti, L.",
      title          = "{The Ideal relativistic rotating gas as a perfect fluid
                        with spin}",
      journal        = "Annals Phys.",
      volume         = "325",
      year           = "2010",
      pages          = "1566-1594",
      doi            = "10.1016/j.aop.2010.03.007",
      eprint         = "0911.0864",
      archivePrefix  = "arXiv",
      primaryClass   = "gr-qc",
      SLACcitation   = "%%CITATION = ARXIV:0911.0864;%%"
}

@article{Becattini:2012tc,
      author         = "Becattini, F.",
      title          = "{Covariant statistical mechanics and the stress-energy
                        tensor}",
      journal        = "Phys. Rev. Lett.",
      volume         = "108",
      year           = "2012",
      pages          = "244502",
      doi            = "10.1103/PhysRevLett.108.244502",
      eprint         = "1201.5278",
      archivePrefix  = "arXiv",
      primaryClass   = "gr-qc",
      SLACcitation   = "%%CITATION = ARXIV:1201.5278;%%"
}

@book{rosenfeld1940tenseur,
  title={Sur le tenseur d'impulsion-{\'e}nergie},
  author={Rosenfeld, L{\'e}on Jacques Henri Constant},
  year={1940},
  publisher={Palais des acad{\'e}mies}
}

@article{BELINFANTE1939887,
title = {On the spin angular momentum of mesons},
journal = {Physica},
volume = {6},
number = {7},
pages = {887-898},
year = {1939},
issn = {0031-8914},
doi = {https://doi.org/10.1016/S0031-8914(39)90090-X},
url = {https://www.sciencedirect.com/science/article/pii/S003189143990090X},
author = {F.J. Belinfante}
}

@article{Belinfante:1940,
title = "On the current and the density of the electric charge, the energy, the linear momentum and the angular momentum of arbitrary fields",
journal = "Physica",
volume = "7",
number = "5",
pages = "449 - 474",
year = "1940",
issn = "0031-8914",
doi = "https://doi.org/10.1016/S0031-8914(40)90091-X",
url = "http://www.sciencedirect.com/science/article/pii/S003189144090091X",
author = "F.J. Belinfante"
}

@article{Becattini:2015nva,
      author         = "Becattini, F. and Grossi, E.",
      title          = "{Quantum corrections to the stress-energy tensor in
                        thermodynamic equilibrium with acceleration}",
      journal        = "Phys. Rev.",
      volume         = "D92",
      year           = "2015",
      pages          = "045037",
      doi            = "10.1103/PhysRevD.92.045037",
      eprint         = "1505.07760",
      archivePrefix  = "arXiv",
      primaryClass   = "gr-qc",
      SLACcitation   = "%%CITATION = ARXIV:1505.07760;%%"
}

@article{Montenegro:2017rbu,
      author         = "Montenegro, David and Tinti, Leonardo and Torrieri,
                        Giorgio",
      title          = "{The ideal relativistic fluid limit for a medium with
                        polarization}",
      journal        = "Phys. Rev.",
      volume         = "D96",
      year           = "2017",
      number         = "5",
      pages          = "056012",
      doi            = "10.1103/PhysRevD.96.056012",
      eprint         = "1701.08263",
      archivePrefix  = "arXiv",
      primaryClass   = "hep-th",
      SLACcitation   = "%%CITATION = ARXIV:1701.08263;%%"
}

@article{Hehl:1976vr,
      author         = "Hehl, F. W.",
      title          = "{On the Energy Tensor of Spinning Massive Matter in
                        Classical Field Theory and General Relativity}",
      journal        = "Rept. Math. Phys.",
      volume         = "9",
      year           = "1976",
      pages          = "55-82",
      doi            = "10.1016/0034-4877(76)90016-1",
      SLACcitation   = "%%CITATION = RMHPB,9,55;%%"
}

@article{Becattini:2014yxa,
      author         = "Becattini, F. and Bucciantini, L. and Grossi, E. and
                        Tinti, L.",
      title          = "{Local thermodynamical equilibrium and the beta frame for
                        a quantum relativistic fluid}",
      journal        = "Eur. Phys. J.",
      volume         = "C75",
      year           = "2015",
      number         = "5",
      pages          = "191",
      doi            = "10.1140/epjc/s10052-015-3384-y",
      eprint         = "1403.6265",
      archivePrefix  = "arXiv",
      primaryClass   = "hep-th",
      SLACcitation   = "%%CITATION = ARXIV:1403.6265;%%"
}

@article{Becattini:2018duy,
      author         = "Becattini, F. and Florkowski, Wojciech and Speranza,
                        Enrico",
      title          = "{Spin tensor and its role in non-equilibrium
                        thermodynamics}",
      journal        = "Phys. Lett.",
      volume         = "B789",
      year           = "2019",
      pages          = "419-425",
      doi            = "10.1016/j.physletb.2018.12.016",
      eprint         = "1807.10994",
      archivePrefix  = "arXiv",
      primaryClass   = "hep-th",
      SLACcitation   = "%%CITATION = ARXIV:1807.10994;%%"
}

@article{Florkowski:2018fap,
    author = "Florkowski, Wojciech and Kumar, Avdhesh and Ryblewski, Radoslaw",
    title = "{Relativistic hydrodynamics for spin-polarized fluids}",
    eprint = "1811.04409",
    archivePrefix = "arXiv",
    primaryClass = "nucl-th",
    doi = "10.1016/j.ppnp.2019.07.001",
    journal = "Prog. Part. Nucl. Phys.",
    volume = "108",
    pages = "103709",
    year = "2019"
}

@article{Florkowski:2017ruc,
      author         = "Florkowski, Wojciech and Friman, Bengt and Jaiswal,
                        Amaresh and Speranza, Enrico",
      title          = "{Relativistic fluid dynamics with spin}",
      journal        = "Phys. Rev.",
      volume         = "C97",
      year           = "2018",
      number         = "4",
      pages          = "041901",
      doi            = "10.1103/PhysRevC.97.041901",
      eprint         = "1705.00587",
      archivePrefix  = "arXiv",
      primaryClass   = "nucl-th",
      SLACcitation   = "%%CITATION = ARXIV:1705.00587;%%"
}

@article{Florkowski:2017dyn,
      author         = "Florkowski, Wojciech and Friman, Bengt and Jaiswal,
                        Amaresh and Ryblewski, Radoslaw and Speranza, Enrico",
      title          = "{Spin-dependent distribution functions for relativistic
                        hydrodynamics of spin-1/2 particles}",
      journal        = "Phys. Rev.",
      volume         = "D97",
      year           = "2018",
      number         = "11",
      pages          = "116017",
      doi            = "10.1103/PhysRevD.97.116017",
      eprint         = "1712.07676",
      archivePrefix  = "arXiv",
      primaryClass   = "nucl-th",
      SLACcitation   = "%%CITATION = ARXIV:1712.07676;%%"
}

@book{Romatschke:2017ejr,
    author = "Romatschke, Paul and Romatschke, Ulrike",
    title = "{Relativistic Fluid Dynamics In and Out of Equilibrium}",
    eprint = "1712.05815",
    archivePrefix = "arXiv",
    primaryClass = "nucl-th",
    doi = "10.1017/9781108651998",
    isbn = "978-1-108-48368-1, 978-1-108-75002-8",
    publisher = "Cambridge University Press",
    series = "Cambridge Monographs on Mathematical Physics",
    month = "5",
    year = "2019"
}

@article{Bhadury:2021oat,
    author = "Bhadury, Samapan and Bhatt, Jitesh and Jaiswal, Amaresh and Kumar, Avdhesh",
    title = "{New developments in relativistic fluid dynamics with spin}",
    eprint = "2101.11964",
    archivePrefix = "arXiv",
    primaryClass = "hep-ph",
    doi = "10.1140/epjs/s11734-021-00020-4",
    journal = "Eur. Phys. J. ST",
    volume = "230",
    number = "3",
    pages = "655--672",
    year = "2021"
}

@article{Vilenkin:1980ft,
    author = "Vilenkin, A.",
    title = "{CANCELLATION OF EQUILIBRIUM PARITY VIOLATING CURRENTS}",
    doi = "10.1103/PhysRevD.22.3067",
    journal = "Phys. Rev. D",
    volume = "22",
    pages = "3067--3079",
    year = "1980"
}

@article{Vilenkin:1980zv,
    author = "Vilenkin, A.",
    title = "{QUANTUM FIELD THEORY AT FINITE TEMPERATURE IN A ROTATING SYSTEM}",
    doi = "10.1103/PhysRevD.21.2260",
    journal = "Phys. Rev. D",
    volume = "21",
    pages = "2260--2269",
    year = "1980"
}

@article{Vilenkin:1978hb,
    author = "Vilenkin, A.",
    title = "{Parity Violating Currents in Thermal Radiation}",
    doi = "10.1016/0370-2693(78)90330-1",
    journal = "Phys. Lett. B",
    volume = "80",
    pages = "150--152",
    year = "1978"
}

@article{Buzzegoli:2023yut,
    author = "Buzzegoli, Matteo and Tuchin, Kirill",
    title = "{Electromagnetic radiation at extreme angular velocity}",
    eprint = "2308.10349",
    archivePrefix = "arXiv",
    primaryClass = "hep-ph",
    doi = "10.1007/JHEP12(2023)113",
    journal = "JHEP",
    volume = "12",
    pages = "113",
    year = "2023"
}

@article{Becattini:2017ljh,
    author = "Becattini, F.",
    title = "{Thermodynamic equilibrium with acceleration and the Unruh effect}",
    eprint = "1712.08031",
    archivePrefix = "arXiv",
    primaryClass = "gr-qc",
    doi = "10.1103/PhysRevD.97.085013",
    journal = "Phys. Rev. D",
    volume = "97",
    number = "8",
    pages = "085013",
    year = "2018"
}

@article{Becattini:2016stj,
    author = "Becattini, F.",
    title = "{Thermodynamic equilibrium in relativity: four-temperature, Killing vectors and Lie derivatives}",
    eprint = "1606.06605",
    archivePrefix = "arXiv",
    primaryClass = "gr-qc",
    doi = "10.5506/APhysPolB.47.1819",
    journal = "Acta Phys. Polon. B",
    volume = "47",
    pages = "1819",
    year = "2016"
}

@article{Buzzegoli:2018wpy,
    author = "Buzzegoli, M. and Becattini, F.",
    title = "{General thermodynamic equilibrium with axial chemical potential for the free Dirac field}",
    eprint = "1807.02071",
    archivePrefix = "arXiv",
    primaryClass = "hep-th",
    doi = "10.1007/JHEP12(2018)002",
    journal = "JHEP",
    volume = "12",
    pages = "002",
    year = "2018",
    note = "[Erratum: JHEP 03, 045 (2022)]"
}

@article{Becattini:2019poj,
    author = "Becattini, F. and Rindori, D.",
    title = "{Extensivity, entropy current, area law and Unruh effect}",
    eprint = "1903.05422",
    archivePrefix = "arXiv",
    primaryClass = "hep-th",
    doi = "10.1103/PhysRevD.99.125011",
    journal = "Phys. Rev. D",
    volume = "99",
    number = "12",
    pages = "125011",
    year = "2019"
}

@article{Becattini:2021lfq,
    author = "Becattini, Francesco and Liao, Jinfeng and Lisa, Michael",
    title = "{Strongly Interacting Matter Under Rotation: An Introduction}",
    eprint = "2102.00933",
    archivePrefix = "arXiv",
    primaryClass = "nucl-th",
    doi = "10.1007/978-3-030-71427-7_1",
    journal = "Lect. Notes Phys.",
    volume = "987",
    pages = "1--14",
    year = "2021"
}

@article{Chernodub:2016kxh,
    author = "Chernodub, M. N. and Gongyo, Shinya",
    title = "{Interacting fermions in rotation: chiral symmetry restoration, moment of inertia and thermodynamics}",
    eprint = "1611.02598",
    archivePrefix = "arXiv",
    primaryClass = "hep-th",
    reportNumber = "RIKEN-QHP-255",
    doi = "10.1007/JHEP01(2017)136",
    journal = "JHEP",
    volume = "01",
    pages = "136",
    year = "2017"
}

@article{Chernodub:2020qah,
    author = "Chernodub, M. N.",
    title = "{Inhomogeneous confining-deconfining phases in rotating plasmas}",
    eprint = "2012.04924",
    archivePrefix = "arXiv",
    primaryClass = "hep-ph",
    doi = "10.1103/PhysRevD.103.054027",
    journal = "Phys. Rev. D",
    volume = "103",
    number = "5",
    pages = "054027",
    year = "2021"
}

@article{Kovtun:2022vas,
    author = "Kovtun, Pavel",
    title = "{Temperature in relativistic fluids}",
    eprint = "2210.15605",
    archivePrefix = "arXiv",
    primaryClass = "gr-qc",
    doi = "10.1103/PhysRevD.107.086012",
    journal = "Phys. Rev. D",
    volume = "107",
    number = "8",
    pages = "086012",
    year = "2023"
}

@article{Kovtun:2019hdm,
    author = "Kovtun, Pavel",
    title = "{First-order relativistic hydrodynamics is stable}",
    eprint = "1907.08191",
    archivePrefix = "arXiv",
    primaryClass = "hep-th",
    doi = "10.1007/JHEP10(2019)034",
    journal = "JHEP",
    volume = "10",
    pages = "034",
    year = "2019"
}

@article{Banerjee:2012iz,
    author = "Banerjee, Nabamita and Bhattacharya, Jyotirmoy and Bhattacharyya, Sayantani and Jain, Sachin and Minwalla, Shiraz and Sharma, Tarun",
    title = "{Constraints on Fluid Dynamics from Equilibrium Partition Functions}",
    eprint = "1203.3544",
    archivePrefix = "arXiv",
    primaryClass = "hep-th",
    reportNumber = "TFR-TH-12-05, IPMU12-0037",
    doi = "10.1007/JHEP09(2012)046",
    journal = "JHEP",
    volume = "09",
    pages = "046",
    year = "2012"
}

@article{Bhattacharyya:2007vs,
    author = "Bhattacharyya, Sayantani and Lahiri, Subhaneil and Loganayagam, R. and Minwalla, Shiraz",
    title = "{Large rotating AdS black holes from fluid mechanics}",
    eprint = "0708.1770",
    archivePrefix = "arXiv",
    primaryClass = "hep-th",
    doi = "10.1088/1126-6708/2008/09/054",
    journal = "JHEP",
    volume = "09",
    pages = "054",
    year = "2008"
}

@article{PhysRev.6.239,
  title = {Magnetization by Rotation},
  author = {Barnett, S. J.},
  journal = {Phys. Rev.},
  volume = {6},
  issue = {4},
  pages = {239--270},
  numpages = {0},
  year = {1915},
  month = {Oct},
  publisher = {American Physical Society},
  doi = {10.1103/PhysRev.6.239},
  url = {https://link.aps.org/doi/10.1103/PhysRev.6.239}
}

@article{PhysRev.10.7,
  title = {The Magnetization of Iron, Nickel, and Cobalt by Rotation and the Nature of the Magnetic Molecule},
  author = {Barnett., S. J.},
  journal = {Phys. Rev.},
  volume = {10},
  issue = {1},
  pages = {7--21},
  numpages = {0},
  year = {1917},
  month = {Jul},
  publisher = {American Physical Society},
  doi = {10.1103/PhysRev.10.7},
  url = {https://link.aps.org/doi/10.1103/PhysRev.10.7}
}

@article{PhysRevLett.122.177202,
  title = {Observation of the Nuclear Barnett Effect},
  author = {Arabgol, Mohsen and Sleator, Tycho},
  journal = {Phys. Rev. Lett.},
  volume = {122},
  issue = {17},
  pages = {177202},
  numpages = {5},
  year = {2019},
  month = {May},
  publisher = {American Physical Society},
  doi = {10.1103/PhysRevLett.122.177202},
  url = {https://link.aps.org/doi/10.1103/PhysRevLett.122.177202}
}

@article{PhysRevResearch.2.033013,
  title = {Exceeding the Landau speed limit with topological Bogoliubov Fermi surfaces},
  author = {Autti, S. and M\"akinen, J. T. and Rysti, J. and Volovik, G. E. and Zavjalov, V. V. and Eltsov, V. B.},
  journal = {Phys. Rev. Res.},
  volume = {2},
  issue = {3},
  pages = {033013},
  numpages = {8},
  year = {2020},
  month = {Jul},
  publisher = {American Physical Society},
  doi = {10.1103/PhysRevResearch.2.033013},
  url = {https://link.aps.org/doi/10.1103/PhysRevResearch.2.033013}
}

@article{Landsman:1986uw,
    author = "Landsman, N. P. and van Weert, C. G.",
    title = "{Real and Imaginary Time Field Theory at Finite Temperature and Density}",
    reportNumber = "ITFA-86-12",
    doi = "10.1016/0370-1573(87)90121-9",
    journal = "Phys. Rept.",
    volume = "145",
    pages = "141",
    year = "1987"
}

@article{green1948general,
  title={A general kinetic theory of liquids V. Liquid He II},
  author={Green, HS},
  journal={Proceedings of the Royal Society of London. Series A. Mathematical and Physical Sciences},
  volume={194},
  number={1037},
  pages={244--258},
  year={1948},
  publisher={The Royal Society London}
}

@article{born1947general,
  title={A general kinetic theory of liquids. IV. Quantum mechanics of fluids},
  author={Born, Max and Green, HS},
  journal={Proceedings of the Royal Society of London. Series A. Mathematical and Physical Sciences},
  volume={191},
  number={1025},
  pages={168--181},
  year={1947},
  publisher={The Royal Society London}
}

@book{Landau_Physical_kinetics,
	author = {Lifshitz, E.M and Pitaevskii, L.P and Pitaevskii, L.P},
	title = {Physical kinetics},
	publisher = {Butterworth-Heinemann},
	year = {1998},
	series = {Landau \& lifshitz course of theoretical physics 10},
	address = {Oxford}
}

@book{Huang1987StatisticalMechanics,
  title     = {Statistical Mechanics},
  author    = {Huang, Kerson},
  edition   = {2nd},
  year      = {1987},
  publisher = {John Wiley \& Sons},
  address   = {New York},
  isbn      = {978-0471815181}
}

@misc{landau1984statistical,
  title={Statistical Physics, Part 1, Butterworth},
  author={Landau, LD and Lifshitz, EM and Reichl, LE},
  year={1984},
  publisher={Heinemann, Oxford, UK}
}

@article{marc2007virial,
  title={The virial theorem},
  author={Marc, Guilhem and McMillan, WG},
  journal={Advances in chemical physics},
  volume={58},
  pages={209--361},
  year={2007}
}

@article{Lin:2015cia,
    author = "Lin, Chris L. and Ordonez, Carlos R.",
    title = "{Virial Theorem for Nonrelativistic Quantum Fields in D Spatial Dimensions}",
    eprint = "1503.05843",
    archivePrefix = "arXiv",
    primaryClass = "hep-th",
    doi = "10.1155/2015/796275",
    journal = "Adv. High Energy Phys.",
    volume = "2015",
    pages = "796275",
    year = "2015"
}

@article{Toyoda:1998ub,
    author = "Toyoda, T. and Takiuchi, K.",
    title = "{Quantum field theoretical reformulation of the virial theorem}",
    journal = "Physica A",
    volume = "261",
    pages = "471--481",
    year = "1998"
}

@article{Ordonez:2015vaa,
    author = "Ordonez, Carlos R.",
    title = "{Path-integral Fujikawa\textquoteright{}s approach to anomalous virial theorems and equations of state for systems with SO(2,1) symmetry}",
    eprint = "1503.01384",
    archivePrefix = "arXiv",
    primaryClass = "cond-mat.quant-gas",
    doi = "10.1016/j.physa.2015.11.019",
    journal = "Physica A",
    volume = "446",
    pages = "64--74",
    year = "2016"
}

@book{de2018superconductivity,
  title={Superconductivity of metals and alloys},
  author={De Gennes, Pierre-Gilles},
  year={2018},
  publisher={CRC press}
}

@article{Fukushima:2018grm,
    author = "Fukushima, Kenji",
    title = "{Extreme matter in electromagnetic fields and rotation}",
    eprint = "1812.08886",
    archivePrefix = "arXiv",
    primaryClass = "hep-ph",
    doi = "10.1016/j.ppnp.2019.04.001",
    journal = "Prog. Part. Nucl. Phys.",
    volume = "107",
    pages = "167--199",
    year = "2019"
}

@article{Becattini:2020ngo,
    author = "Becattini, Francesco and Lisa, Michael A.",
    title = "{Polarization and Vorticity in the Quark\textendash{}Gluon Plasma}",
    eprint = "2003.03640",
    archivePrefix = "arXiv",
    primaryClass = "nucl-ex",
    doi = "10.1146/annurev-nucl-021920-095245",
    journal = "Ann. Rev. Nucl. Part. Sci.",
    volume = "70",
    pages = "395--423",
    year = "2020"
}

@article{Huang:2020dtn,
    author = "Huang, Xu-Guang and Liao, Jinfeng and Wang, Qun and Xia, Xiao-Liang",
    title = "{Vorticity and Spin Polarization in Heavy Ion Collisions: Transport Models}",
    eprint = "2010.08937",
    archivePrefix = "arXiv",
    primaryClass = "nucl-th",
    doi = "10.1007/978-3-030-71427-7_9",
    journal = "Lect. Notes Phys.",
    volume = "987",
    pages = "281--308",
    year = "2021"
}

@article{Wang:2017jpl,
    author = "Wang, Qun",
    editor = "Heinz, Ulrich and Evdokimov, Olga and Jacobs, Peter",
    title = "{Global and local spin polarization in heavy ion collisions: a brief overview}",
    eprint = "1704.04022",
    archivePrefix = "arXiv",
    primaryClass = "nucl-th",
    doi = "10.1016/j.nuclphysa.2017.06.053",
    journal = "Nucl. Phys. A",
    volume = "967",
    pages = "225--232",
    year = "2017"
}

@article{Dey:2024crk,
    author = "Dey, Sourav and Bhadury, Samapan and Florkowski, Wojciech and Ryblewski, Radoslaw and Jaiswal, Amaresh",
    title = "{Energy-momentum correlators of fermions at finite temperature and density}",
    eprint = "2409.18912",
    archivePrefix = "arXiv",
    primaryClass = "hep-ph",
    doi = "10.1103/PhysRevD.110.116002",
    journal = "Phys. Rev. D",
    volume = "110",
    number = "11",
    pages = "116002",
    year = "2024"
}

@article{Becattini:2022zvf,
    author = "Becattini, Francesco",
    title = "{Spin and polarization: a new direction in relativistic heavy ion physics}",
    eprint = "2204.01144",
    archivePrefix = "arXiv",
    primaryClass = "nucl-th",
    doi = "10.1088/1361-6633/ac97a9",
    journal = "Rept. Prog. Phys.",
    volume = "85",
    number = "12",
    pages = "122301",
    year = "2022"
}

@article{Chen:2015hfc,
    author = "Chen, Hao-Lei and Fukushima, Kenji and Huang, Xu-Guang and Mameda, Kazuya",
    title = "{Analogy between rotation and density for Dirac fermions in a magnetic field}",
    eprint = "1512.08974",
    archivePrefix = "arXiv",
    primaryClass = "hep-ph",
    doi = "10.1103/PhysRevD.93.104052",
    journal = "Phys. Rev. D",
    volume = "93",
    number = "10",
    pages = "104052",
    year = "2016"
}

@article{Zhang:2020hha,
    author = "Zhang, Zheng and Shi, Chao and He, Xiao-Tao and Luo, Xiaofeng and Zong, Hong-Shi",
    title = "{Chiral phase transition inside a rotating cylinder within the Nambu\textendash{}Jona-Lasinio model}",
    eprint = "2012.01017",
    archivePrefix = "arXiv",
    primaryClass = "hep-ph",
    doi = "10.1103/PhysRevD.102.114023",
    journal = "Phys. Rev. D",
    volume = "102",
    number = "11",
    pages = "114023",
    year = "2020"
}

@article{PhysRevD.102.114023,
  title = {Chiral phase transition inside a rotating cylinder within the Nambu--Jona-Lasinio model},
  author = {Zhang, Zheng and Shi, Chao and He, Xiao-Tao and Luo, Xiaofeng and Zong, Hong-Shi},
  journal = {Phys. Rev. D},
  volume = {102},
  issue = {11},
  pages = {114023},
  numpages = {8},
  year = {2020},
  month = {Dec},
  publisher = {American Physical Society},
  doi = {10.1103/PhysRevD.102.114023},
  url = {https://link.aps.org/doi/10.1103/PhysRevD.102.114023}
}

@article{PhysRevD.93.104052,
  title = {Analogy between rotation and density for Dirac fermions in a magnetic field},
  author = {Chen, Hao-Lei and Fukushima, Kenji and Huang, Xu-Guang and Mameda, Kazuya},
  journal = {Phys. Rev. D},
  volume = {93},
  issue = {10},
  pages = {104052},
  numpages = {12},
  year = {2016},
  month = {May},
  publisher = {American Physical Society},
  doi = {10.1103/PhysRevD.93.104052},
  url = {https://link.aps.org/doi/10.1103/PhysRevD.93.104052}
}

@article{PhysRevLett.117.192302,
  title = {Pairing Phase Transitions of Matter under Rotation},
  author = {Jiang, Yin and Liao, Jinfeng},
  journal = {Phys. Rev. Lett.},
  volume = {117},
  issue = {19},
  pages = {192302},
  numpages = {5},
  year = {2016},
  month = {Nov},
  publisher = {American Physical Society},
  doi = {10.1103/PhysRevLett.117.192302},
  url = {https://link.aps.org/doi/10.1103/PhysRevLett.117.192302}
}

@article{PhysRevD.95.096006,
  title = {Effects of rotation and boundaries on chiral symmetry breaking of relativistic fermions},
  author = {Chernodub, M. N. and Gongyo, Shinya},
  journal = {Phys. Rev. D},
  volume = {95},
  issue = {9},
  pages = {096006},
  numpages = {10},
  year = {2017},
  month = {May},
  publisher = {American Physical Society},
  doi = {10.1103/PhysRevD.95.096006},
  url = {https://link.aps.org/doi/10.1103/PhysRevD.95.096006}
}

@article{Chen:2020ath,
    author = "Chen, Xun and Zhang, Lin and Li, Danning and Hou, Defu and Huang, Mei",
    title = "{Gluodynamics and deconfinement phase transition under rotation from holography}",
    eprint = "2010.14478",
    archivePrefix = "arXiv",
    primaryClass = "hep-ph",
    doi = "10.1007/JHEP07(2021)132",
    journal = "JHEP",
    volume = "07",
    pages = "132",
    year = "2021"
}

@article{PhysRevD.107.114502,
  title = {Inhomogeneity of a rotating gluon plasma and the Tolman-Ehrenfest law in imaginary time: Lattice results for fast imaginary rotation},
  author = {Chernodub, M. N. and Goy, V. A. and Molochkov, A. V.},
  journal = {Phys. Rev. D},
  volume = {107},
  issue = {11},
  pages = {114502},
  numpages = {14},
  year = {2023},
  month = {Jun},
  publisher = {American Physical Society},
  doi = {10.1103/PhysRevD.107.114502},
  url = {https://link.aps.org/doi/10.1103/PhysRevD.107.114502}
}

@article{Braguta:2020biu,
    author = "Braguta, V. V. and Kotov, A. Yu. and Kuznedelev, D. D. and Roenko, A. A.",
    title = "{Study of the Confinement/Deconfinement Phase Transition in Rotating Lattice SU(3) Gluodynamics}",
    doi = "10.31857/S1234567820130029",
    journal = "Pisma Zh. Eksp. Teor. Fiz.",
    volume = "112",
    number = "1",
    pages = "9--16",
    year = "2020"
}

@article{Fujimoto:2021xix,
    author = "Fujimoto, Yuki and Fukushima, Kenji and Hidaka, Yoshimasa",
    title = "{Deconfining Phase Boundary of Rapidly Rotating Hot and Dense Matter and Analysis of Moment of Inertia}",
    eprint = "2101.09173",
    archivePrefix = "arXiv",
    primaryClass = "hep-ph",
    reportNumber = "KEK-TH-2290, J-PARC-TH-0236, RIKEN-iTHEMS-Report-21",
    doi = "10.1016/j.physletb.2021.136184",
    journal = "Phys. Lett. B",
    volume = "816",
    pages = "136184",
    year = "2021"
}

@article{PhysRevD.103.094515,
  title = {Influence of relativistic rotation on the confinement-deconfinement transition in gluodynamics},
  author = {Braguta, V. V. and Kotov, A. Yu. and Kuznedelev, D. D. and Roenko, A. A.},
  journal = {Phys. Rev. D},
  volume = {103},
  issue = {9},
  pages = {094515},
  numpages = {16},
  year = {2021},
  month = {May},
  publisher = {American Physical Society},
  doi = {10.1103/PhysRevD.103.094515},
  url = {https://link.aps.org/doi/10.1103/PhysRevD.103.094515}
}

@article{Callan:1970ze,
    author = "Callan, Jr., Curtis G. and Coleman, Sidney R. and Jackiw, Roman",
    title = "{A New improved energy - momentum tensor}",
    doi = "10.1016/0003-4916(70)90394-5",
    journal = "Annals Phys.",
    volume = "59",
    pages = "42--73",
    year = "1970"
}

@article{Coleman:1970je,
    author = "Coleman, Sidney R. and Jackiw, Roman",
    title = "{Why dilatation generators do not generate dilatations?}",
    doi = "10.1016/0003-4916(71)90153-9",
    journal = "Annals Phys.",
    volume = "67",
    pages = "552--598",
    year = "1971"
}

@article{PhysRevD.2.1541,
  title = {Broken Scale Invariance in Scalar Field Theory},
  author = {Callan, Curtis G.},
  journal = {Phys. Rev. D},
  volume = {2},
  issue = {8},
  pages = {1541--1547},
  numpages = {0},
  year = {1970},
  month = {Oct},
  publisher = {American Physical Society},
  doi = {10.1103/PhysRevD.2.1541},
  url = {https://link.aps.org/doi/10.1103/PhysRevD.2.1541}
}

@article{Jackiw:1971dp,
    author = "Jackiw, R.",
    editor = "Barut, A. O. and Brittin, W. E.",
    title = "{A NEW IMPROVED ENERGY MOMENTUM TENSOR.}",
    reportNumber = "MIT-CTP-144",
    journal = "Lect. Theor. Phys.",
    volume = "13",
    pages = "299--302",
    year = "1971"
}

@article{Jackiw:1968lmf,
    author = "Jackiw, R.",
    title = "{Stress-tensor-current commutators, electromagnetic and weak corrections to current commutators, and sum rules}",
    doi = "10.1103/PhysRev.175.2058",
    journal = "Phys. Rev.",
    volume = "175",
    pages = "2058--2065",
    year = "1968"
}

@article{Genz:1970gzi,
    author = "Genz, H. and Katz, J.",
    title = "{Derivation of equal time commutators involving the symmetric energy-momentum tensor and applications}",
    doi = "10.1103/PhysRevD.2.2225",
    journal = "Phys. Rev. D",
    volume = "2",
    pages = "2225--2240",
    year = "1970"
}

@article{Banks:1972jf,
    author = "Banks, Tom",
    title = "{Ward identities for the energy-momentum tensor}",
    doi = "10.1103/PhysRevD.5.2959",
    journal = "Phys. Rev. D",
    volume = "5",
    pages = "2959--2965",
    year = "1972"
}

@article{Gross:1967zz,
    author = "Gross, David J. and Jackiw, Roman",
    title = "{Derivation of the SU (3) X SU (3) Space-Time Local Current Commutators}",
    doi = "10.1103/PhysRev.163.1688",
    journal = "Phys. Rev.",
    volume = "163",
    pages = "1688--1696",
    year = "1967"
}

@article{Beg:1970ai,
    author = "Beg, M. A. B. and Bernstein, J. and Gross, D. J. and Jackiw, R. and Sirlin, A.",
    title = "{Dimensions of currents and current commutators}",
    doi = "10.1103/PhysRevLett.25.1231",
    journal = "Phys. Rev. Lett.",
    volume = "25",
    pages = "1231--1234",
    year = "1970"
}

@article{Levin:1971sw,
    author = "Levin, David N.",
    title = "{Formal restrictions on schwinger terms in commutators containing j(mu) and/or theta(mu nu)}",
    reportNumber = "PRINT-70-2772",
    doi = "10.1103/PhysRevD.3.1320",
    journal = "Phys. Rev. D",
    volume = "3",
    pages = "1320--1326",
    year = "1971"
}

@article{Deser:1967zzf,
    author = "Deser, Stanley and Boulware, D.",
    title = "{Stress-Tensor Commutators and Schwinger Terms}",
    doi = "10.1063/1.1705368",
    journal = "J. Math. Phys.",
    volume = "8",
    pages = "1468",
    year = "1967"
}

@article{Schwinger:1963zz,
    author = "Schwinger, Julian",
    title = "{Energy and Momentum Density in Field Theory}",
    doi = "10.1103/PhysRev.130.800",
    journal = "Phys. Rev.",
    volume = "130",
    pages = "800--805",
    year = "1963"
}

@article{Trubatch:1970zr,
    author = "Trubatch, J.",
    title = "{Stress-tensor commutation relations}",
    doi = "10.1007/BF02754155",
    journal = "Nuovo Cim. A",
    volume = "68",
    pages = "339--367",
    year = "1970"
}

@book{francesco2012conformal,
  title={Conformal field theory},
  author={Francesco, Philippe and Mathieu, Pierre and S{\'e}n{\'e}chal, David},
  year={2012},
  publisher={Springer Science \& Business Media}
}

@article{Becattini:2020sww,
    author = "Becattini, Francesco",
    title = "{Polarization in Relativistic Fluids: A Quantum Field Theoretical Derivation}",
    eprint = "2004.04050",
    archivePrefix = "arXiv",
    primaryClass = "hep-th",
    doi = "10.1007/978-3-030-71427-7_2",
    journal = "Lect. Notes Phys.",
    volume = "987",
    pages = "15--52",
    year = "2021"
}

@article{Ambrus:2019ayb,
    author = "Ambrus, Victor E.",
    title = "{Helical massive fermions under rotation}",
    eprint = "1912.09977",
    archivePrefix = "arXiv",
    primaryClass = "nucl-th",
    doi = "10.1007/JHEP08(2020)016",
    journal = "JHEP",
    volume = "08",
    pages = "016",
    year = "2020"
}

@article{Ambrus:2019khr,
    author = "Ambrus, Victor E. and Chernodub, M. N.",
    title = "{Vortical effects in Dirac fluids with vector, chiral and helical charges}",
    eprint = "1912.11034",
    archivePrefix = "arXiv",
    primaryClass = "hep-th",
    doi = "10.1140/epjc/s10052-023-11244-0",
    journal = "Eur. Phys. J. C",
    volume = "83",
    number = "2",
    pages = "111",
    year = "2023",
    note = "[Erratum: Eur.Phys.J.C 84, 289 (2024)]"
}

@article{Kovtun:2012rj,
    author = "Kovtun, Pavel",
    title = "{Lectures on hydrodynamic fluctuations in relativistic theories}",
    eprint = "1205.5040",
    archivePrefix = "arXiv",
    primaryClass = "hep-th",
    doi = "10.1088/1751-8113/45/47/473001",
    journal = "J. Phys. A",
    volume = "45",
    pages = "473001",
    year = "2012"
}

@article{weinhold1975metric,
  title={Metric geometry of equilibrium thermodynamics},
  author={Weinhold, Frank},
  journal={The Journal of Chemical Physics},
  volume={63},
  number={6},
  pages={2479--2483},
  year={1975},
  publisher={AIP Publishing}
}

@article{Braguta:2023tqz,
    author = "Braguta, Victor V. and Chernodub, Maxim N. and Kudrov, Ilya E. and Roenko, Artem A. and Sychev, Dmitrii A.",
    title = "{Negative Barnett effect, negative moment of inertia of the gluon plasma, and thermal evaporation of the chromomagnetic condensate}",
    eprint = "2310.16036",
    archivePrefix = "arXiv",
    primaryClass = "hep-ph",
    doi = "10.1103/PhysRevD.110.014511",
    journal = "Phys. Rev. D",
    volume = "110",
    number = "1",
    pages = "014511",
    year = "2024"
}

@article{Braguta:2023yjn,
    author = "Braguta, Victor V. and Chernodub, Maxim N. and Roenko, Artem A. and Sychev, Dmitrii A.",
    title = "{Negative moment of inertia and rotational instability of gluon plasma}",
    eprint = "2303.03147",
    archivePrefix = "arXiv",
    primaryClass = "hep-lat",
    doi = "10.1016/j.physletb.2024.138604",
    journal = "Phys. Lett. B",
    volume = "852",
    pages = "138604",
    year = "2024"
}

@article{Braguta:2023kwl,
    author = "Braguta, V. V. and Kudrov, I. E. and Roenko, A. A. and Sychev, D. A. and Chernodub, M. N.",
    title = "{Lattice Study of the Equation of State of a Rotating Gluon Plasma}",
    doi = "10.1134/S0021364023600830",
    journal = "JETP Lett.",
    volume = "117",
    number = "9",
    pages = "639--644",
    year = "2023"
}

@article{Hata:1980yr,
    author = "Hata, Hiroyuki and Kugo, Taichiro",
    title = "{An Operator Formalism of Statistical Mechanics of Gauge Theory in Covariant Gauges}",
    reportNumber = "KUNS-527",
    doi = "10.1103/PhysRevD.21.3333",
    journal = "Phys. Rev. D",
    volume = "21",
    pages = "3333",
    year = "1980"
}

@article{PhysRevD.16.1130,
  title = {Fermions and gauge vector mesons at finite temperature and density. I. Formal techniques},
  author = {Freedman, Barry A. and McLerran, Larry D.},
  journal = {Phys. Rev. D},
  volume = {16},
  issue = {4},
  pages = {1130--1146},
  numpages = {0},
  year = {1977},
  month = {Aug},
  publisher = {American Physical Society},
  doi = {10.1103/PhysRevD.16.1130},
  url = {https://link.aps.org/doi/10.1103/PhysRevD.16.1130}
}

@article{Freedman:1976dm,
    author = "Freedman, Barry A. and McLerran, Larry D.",
    title = "{Fermions and Gauge Vector Mesons at Finite Temperature and Density. 2. The Ground State Energy of a Relativistic electron Gas}",
    reportNumber = "MIT-CTP-576",
    doi = "10.1103/PhysRevD.16.1147",
    journal = "Phys. Rev. D",
    volume = "16",
    pages = "1147",
    year = "1977"
}

@article{Ojima:1981ma,
    author = "Ojima, Izumi",
    title = "{Gauge Fields at Finite Temperatures: Thermo Field Dynamics, KMS Condition and their Extension to Gauge Theories}",
    reportNumber = "Print-81-0302 (IAS, PRINCETON)",
    doi = "10.1016/0003-4916(81)90058-0",
    journal = "Annals Phys.",
    volume = "137",
    pages = "1",
    year = "1981"
}

@article{Freedman:1974gs,
    author = "Freedman, Daniel Z. and Muzinich, Ivan J. and Weinberg, Erick J.",
    title = "{On the Energy-Momentum Tensor in Gauge Field Theories}",
    reportNumber = "BNL-18740, COO-2220-18",
    doi = "10.1016/0003-4916(74)90448-5",
    journal = "Annals Phys.",
    volume = "87",
    pages = "95",
    year = "1974"
}

@book{Pascual:1984zb,
    author = "Pascual, P. and Tarrach, R.",
    title = "{QCD: RENORMALIZATION FOR THE PRACTITIONER}",
    volume = "194",
    year = "1984"
}

@article{PhysRevB.104.184414,
  title = {Hydrodynamic theory of vorticity-induced spin transport},
  author = {Tatara, Gen},
  journal = {Phys. Rev. B},
  volume = {104},
  issue = {18},
  pages = {184414},
  numpages = {13},
  year = {2021},
  month = {Nov},
  publisher = {American Physical Society},
  doi = {10.1103/PhysRevB.104.184414},
  url = {https://link.aps.org/doi/10.1103/PhysRevB.104.184414}
}

@article{PhysRevB.96.020401,
  title = {Theory of spin hydrodynamic generation},
  author = {Matsuo, M. and Ohnuma, Y. and Maekawa, S.},
  journal = {Phys. Rev. B},
  volume = {96},
  issue = {2},
  pages = {020401},
  numpages = {5},
  year = {2017},
  month = {Jul},
  publisher = {American Physical Society},
  doi = {10.1103/PhysRevB.96.020401},
  url = {https://link.aps.org/doi/10.1103/PhysRevB.96.020401}
}

@article{takahashi2016spin,
  title={Spin hydrodynamic generation},
  author={Takahashi, R and Matsuo, M and Ono, M and Harii, K and Chudo, H and Okayasu, S and Ieda, J and Takahashi, S and Maekawa, S and Saitoh, E},
  journal={Nature Physics},
  volume={12},
  number={1},
  pages={52--56},
  year={2016},
  publisher={Nature Publishing Group UK London}
}

@article{PhysRevLett.106.076601,
  title = {Effects of Mechanical Rotation on Spin Currents},
  author = {Matsuo, Mamoru and Ieda, Jun'ichi and Saitoh, Eiji and Maekawa, Sadamichi},
  journal = {Phys. Rev. Lett.},
  volume = {106},
  issue = {7},
  pages = {076601},
  numpages = {4},
  year = {2011},
  month = {Feb},
  publisher = {American Physical Society},
  doi = {10.1103/PhysRevLett.106.076601},
  url = {https://link.aps.org/doi/10.1103/PhysRevLett.106.076601}
}

@article{Weickgenannt:2024esg,
    author = "Weickgenannt, Nora and Blaizot, Jean-Paul",
    title = "{Spin polarization of an expanding and rotating system}",
    eprint = "2412.05733",
    archivePrefix = "arXiv",
    primaryClass = "nucl-th",
    doi = "10.1103/PhysRevD.111.076007",
    journal = "Phys. Rev. D",
    volume = "111",
    number = "7",
    pages = "076007",
    year = "2025"
}

@article{Sheng:2021kfc,
    author = "Sheng, Xin-Li and Weickgenannt, Nora and Speranza, Enrico and Rischke, Dirk H. and Wang, Qun",
    title = "{From Kadanoff-Baym to Boltzmann equations for massive spin-1/2 fermions}",
    eprint = "2103.10636",
    archivePrefix = "arXiv",
    primaryClass = "nucl-th",
    reportNumber = "USTC-ICTS/PCFT-21-12",
    doi = "10.1103/PhysRevD.104.016029",
    journal = "Phys. Rev. D",
    volume = "104",
    number = "1",
    pages = "016029",
    year = "2021"
}

@article{Bhadury:2022ulr,
    author = "Bhadury, Samapan and Florkowski, Wojciech and Jaiswal, Amaresh and Kumar, Avdhesh and Ryblewski, Radoslaw",
    title = "{Relativistic Spin Magnetohydrodynamics}",
    eprint = "2204.01357",
    archivePrefix = "arXiv",
    primaryClass = "nucl-th",
    doi = "10.1103/PhysRevLett.129.192301",
    journal = "Phys. Rev. Lett.",
    volume = "129",
    number = "19",
    pages = "192301",
    year = "2022"
}

@article{Bhadury:2020puc,
    author = "Bhadury, Samapan and Florkowski, Wojciech and Jaiswal, Amaresh and Kumar, Avdhesh and Ryblewski, Radoslaw",
    title = "{Relativistic dissipative spin dynamics in the relaxation time approximation}",
    eprint = "2002.03937",
    archivePrefix = "arXiv",
    primaryClass = "hep-ph",
    doi = "10.1016/j.physletb.2021.136096",
    journal = "Phys. Lett. B",
    volume = "814",
    pages = "136096",
    year = "2021"
}

@article{Bhadury:2020cop,
    author = "Bhadury, Samapan and Florkowski, Wojciech and Jaiswal, Amaresh and Kumar, Avdhesh and Ryblewski, Radoslaw",
    title = "{Dissipative Spin Dynamics in Relativistic Matter}",
    eprint = "2008.10976",
    archivePrefix = "arXiv",
    primaryClass = "nucl-th",
    doi = "10.1103/PhysRevD.103.014030",
    journal = "Phys. Rev. D",
    volume = "103",
    number = "1",
    pages = "014030",
    year = "2021"
}

@article{Lucas:2017idv,
    author = "Lucas, Andrew and Fong, Kin Chung",
    title = "{Hydrodynamics of electrons in graphene}",
    eprint = "1710.08425",
    archivePrefix = "arXiv",
    primaryClass = "cond-mat.str-el",
    doi = "10.1088/1361-648X/aaa274",
    journal = "J. Phys. Condens. Matter",
    volume = "30",
    number = "5",
    pages = "053001",
    year = "2018"
}

@article{Jaiswal:2024urq,
    author        = "Jaiswal, Amaresh",
    title         = "{Spin-hydrodynamics of electrons in graphene and magnetization due to thermal vorticity}",
    eprint        = "2409.07764",
    archivePrefix = "arXiv",
    primaryClass  = "cond-mat.mes-hall",
    month         = "9",
    year          = "2024",
    note          = "arXiv:2409.07764 [cond-mat.mes-hall]"
}

@article{Braguta:2023iyx,
    author = "Braguta, Victor V. and Chernodub, Maxim N. and Roenko, Artem A.",
    title = "{New mixed inhomogeneous phase in vortical gluon plasma: First-principle results from rotating SU(3) lattice gauge theory}",
    eprint = "2312.13994",
    archivePrefix = "arXiv",
    primaryClass = "hep-lat",
    doi = "10.1016/j.physletb.2024.138783",
    journal = "Phys. Lett. B",
    volume = "855",
    pages = "138783",
    year = "2024"
}

@article{Watts:2016uzu,
    author = "Watts, Anna L. and others",
    title = "{Colloquium : Measuring the neutron star equation of state using x-ray timing}",
    eprint = "1602.01081",
    archivePrefix = "arXiv",
    primaryClass = "astro-ph.HE",
    doi = "10.1103/RevModPhys.88.021001",
    journal = "Rev. Mod. Phys.",
    volume = "88",
    number = "2",
    pages = "021001",
    year = "2016"
}

@article{Grenier:2015pya,
    author = "Grenier, Isabelle A. and Harding, Alice K.",
    title = "{Gamma-ray pulsars: a gold mine}",
    eprint = "1509.08823",
    archivePrefix = "arXiv",
    primaryClass = "astro-ph.HE",
    doi = "10.1016/j.crhy.2015.08.013",
    journal = "Comptes Rendus Physique",
    volume = "16",
    pages = "641--660",
    year = "2015"
}

@article{Basar:2013iaa,
    author = "Basar, Gokce and Kharzeev, Dmitri E. and Yee, Ho-Ung",
    title = "{Triangle anomaly in Weyl semimetals}",
    eprint = "1305.6338",
    archivePrefix = "arXiv",
    primaryClass = "hep-th",
    doi = "10.1103/PhysRevB.89.035142",
    journal = "Phys. Rev. B",
    volume = "89",
    number = "3",
    pages = "035142",
    year = "2014"
}

@article{Landsteiner:2013sja,
    author = "Landsteiner, Karl",
    title = "{Anomalous transport of Weyl fermions in Weyl semimetals}",
    eprint = "1306.4932",
    archivePrefix = "arXiv",
    primaryClass = "hep-th",
    reportNumber = "IFT-UAM-CSIC-13-074, INT-PUB-13-022",
    doi = "10.1103/PhysRevB.89.075124",
    journal = "Phys. Rev. B",
    volume = "89",
    number = "7",
    pages = "075124",
    year = "2014"
}

@article{STAR:2007ccu,
    author = "Abelev, B. I. and others",
    collaboration = "STAR",
    title = "{Global polarization measurement in Au+Au collisions}",
    eprint = "0705.1691",
    archivePrefix = "arXiv",
    primaryClass = "nucl-ex",
    reportNumber = "STAR-05-11-2007",
    doi = "10.1103/PhysRevC.76.024915",
    journal = "Phys. Rev. C",
    volume = "76",
    pages = "024915",
    year = "2007",
    note = "[Erratum: Phys.Rev.C 95, 039906 (2017)]"
}

@article{Heinz:2013th,
    author = "Heinz, Ulrich and Snellings, Raimond",
    title = "{Collective flow and viscosity in relativistic heavy-ion collisions}",
    eprint = "1301.2826",
    archivePrefix = "arXiv",
    primaryClass = "nucl-th",
    doi = "10.1146/annurev-nucl-102212-170540",
    journal = "Ann. Rev. Nucl. Part. Sci.",
    volume = "63",
    pages = "123--151",
    year = "2013"
}

@article{Romatschke:2007mq,
    author = "Romatschke, Paul and Romatschke, Ulrike",
    title = "{Viscosity Information from Relativistic Nuclear Collisions: How Perfect is the Fluid Observed at RHIC?}",
    eprint = "0706.1522",
    archivePrefix = "arXiv",
    primaryClass = "nucl-th",
    reportNumber = "INT-PUB-07-14",
    doi = "10.1103/PhysRevLett.99.172301",
    journal = "Phys. Rev. Lett.",
    volume = "99",
    pages = "172301",
    year = "2007"
}

@article{Jiang:2016wvv,
    author = "Jiang, Yin and Liao, Jinfeng",
    title = "{Pairing Phase Transitions of Matter under Rotation}",
    eprint = "1606.03808",
    archivePrefix = "arXiv",
    primaryClass = "hep-ph",
    doi = "10.1103/PhysRevLett.117.192302",
    journal = "Phys. Rev. Lett.",
    volume = "117",
    number = "19",
    pages = "192302",
    year = "2016"
}

@article{Csernai:2013bqa,
    author = "Csernai, L. P. and Magas, V. K. and Wang, D. J.",
    title = "{Flow Vorticity in Peripheral High Energy Heavy Ion Collisions}",
    eprint = "1302.5310",
    archivePrefix = "arXiv",
    primaryClass = "nucl-th",
    doi = "10.1103/PhysRevC.87.034906",
    journal = "Phys. Rev. C",
    volume = "87",
    number = "3",
    pages = "034906",
    year = "2013"
}

@article{Deng:2016gyh,
    author = "Deng, Wei-Tian and Huang, Xu-Guang",
    title = "{Vorticity in Heavy-Ion Collisions}",
    eprint = "1603.06117",
    archivePrefix = "arXiv",
    primaryClass = "nucl-th",
    doi = "10.1103/PhysRevC.93.064907",
    journal = "Phys. Rev. C",
    volume = "93",
    number = "6",
    pages = "064907",
    year = "2016"
}

@article{Jiang:2016woz,
    author = "Jiang, Yin and Lin, Zi-Wei and Liao, Jinfeng",
    title = "{Rotating quark-gluon plasma in relativistic heavy ion collisions}",
    eprint = "1602.06580",
    archivePrefix = "arXiv",
    primaryClass = "hep-ph",
    doi = "10.1103/PhysRevC.94.044910",
    journal = "Phys. Rev. C",
    volume = "94",
    number = "4",
    pages = "044910",
    year = "2016",
    note = "[Erratum: Phys.Rev.C 95, 049904 (2017)]"
}

@article{Braga:2023qej,
    author = "Braga, Nelson R. F. and Junqueira, Octavio C.",
    title = "{Inhomogeneity of a rotating quark-gluon plasma from holography}",
    eprint = "2306.08653",
    archivePrefix = "arXiv",
    primaryClass = "hep-th",
    doi = "10.1016/j.physletb.2023.138330",
    journal = "Phys. Lett. B",
    volume = "848",
    pages = "138330",
    year = "2024"
}

@article{Zhao:2022uxc,
    author = "Zhao, Yan-Qing and He, Song and Hou, Defu and Li, Li and Li, Zhibin",
    title = "{Phase diagram of holographic thermal dense QCD matter with rotation}",
    eprint = "2212.14662",
    archivePrefix = "arXiv",
    primaryClass = "hep-ph",
    doi = "10.1007/JHEP04(2023)115",
    journal = "JHEP",
    volume = "04",
    pages = "115",
    year = "2023"
}

@article{Chen:2022smf,
    author = "Chen, Shi and Fukushima, Kenji and Shimada, Yusuke",
    title = "{Perturbative Confinement in Thermal Yang-Mills Theories Induced by Imaginary Angular Velocity}",
    eprint = "2207.12665",
    archivePrefix = "arXiv",
    primaryClass = "hep-ph",
    doi = "10.1103/PhysRevLett.129.242002",
    journal = "Phys. Rev. Lett.",
    volume = "129",
    number = "24",
    pages = "242002",
    year = "2022"
}

@article{Sadooghi:2021upd,
    author = "Sadooghi, N. and Tabatabaee Mehr, S. M. A. and Taghinavaz, F.",
    title = "{Inverse magnetorotational catalysis and the phase diagram of a rotating hot and magnetized quark matter}",
    eprint = "2108.12760",
    archivePrefix = "arXiv",
    primaryClass = "hep-ph",
    doi = "10.1103/PhysRevD.104.116022",
    journal = "Phys. Rev. D",
    volume = "104",
    number = "11",
    pages = "116022",
    year = "2021"
}

@article{Wang:2018sur,
    author = "Wang, Xinyang and Wei, Minghua and Li, Zhibin and Huang, Mei",
    title = "{Quark matter under rotation in the NJL model with vector interaction}",
    eprint = "1808.01931",
    archivePrefix = "arXiv",
    primaryClass = "hep-ph",
    doi = "10.1103/PhysRevD.99.016018",
    journal = "Phys. Rev. D",
    volume = "99",
    number = "1",
    pages = "016018",
    year = "2019"
}

@article{Yamamoto:2013zwa,
    author = "Yamamoto, Arata and Hirono, Yuji",
    title = "{Lattice QCD in rotating frames}",
    eprint = "1303.6292",
    archivePrefix = "arXiv",
    primaryClass = "hep-lat",
    reportNumber = "RIKEN-QHP-80",
    doi = "10.1103/PhysRevLett.111.081601",
    journal = "Phys. Rev. Lett.",
    volume = "111",
    pages = "081601",
    year = "2013"
}

@article{Liu:2018kfw,
    author = "Liu, Hong and Glorioso, Paolo",
    title = "{Lectures on non-equilibrium effective field theories and fluctuating hydrodynamics}",
    eprint = "1805.09331",
    archivePrefix = "arXiv",
    primaryClass = "hep-th",
    reportNumber = "MIT-CTP/5018; EFI-18-8, MIT-CTP-5018, EFI-18-8",
    doi = "10.22323/1.305.0008",
    journal = "PoS",
    volume = "TASI2017",
    pages = "008",
    year = "2018"
}

@article{Dey:2023hft,
    author = "Dey, Sourav and Florkowski, Wojciech and Jaiswal, Amaresh and Ryblewski, Radoslaw",
    title = "{Pseudogauge freedom and the SO(3) algebra of spin operators}",
    eprint = "2303.05271",
    archivePrefix = "arXiv",
    primaryClass = "hep-th",
    doi = "10.1016/j.physletb.2023.137994",
    journal = "Phys. Lett. B",
    volume = "843",
    pages = "137994",
    year = "2023"
}

@article{Ayala:2019iin,
    author = "Ayala, Alejandro and De La Cruz, David and Hern\'andez-Ort\'\i{}z, S. and Hern\'andez, L. A. and Salinas, Jordi",
    title = "{Relaxation time for quark spin and thermal vorticity alignment in heavy-ion collisions}",
    eprint = "1909.00274",
    archivePrefix = "arXiv",
    primaryClass = "hep-ph",
    doi = "10.1016/j.physletb.2019.135169",
    journal = "Phys. Lett. B",
    volume = "801",
    pages = "135169",
    year = "2020"
}

@article{Kapusta:2019sad,
    author = "Kapusta, Joseph I. and Rrapaj, Ermal and Rudaz, Serge",
    title = "{Relaxation Time for Strange Quark Spin in Rotating Quark-Gluon Plasma}",
    eprint = "1907.10750",
    archivePrefix = "arXiv",
    primaryClass = "nucl-th",
    doi = "10.1103/PhysRevC.101.024907",
    journal = "Phys. Rev. C",
    volume = "101",
    number = "2",
    pages = "024907",
    year = "2020"
}

@article{Weickgenannt:2020aaf,
    author = "Weickgenannt, Nora and Speranza, Enrico and Sheng, Xin-li and Wang, Qun and Rischke, Dirk H.",
    title = "{Generating Spin Polarization from Vorticity through Nonlocal Collisions}",
    eprint = "2005.01506",
    archivePrefix = "arXiv",
    primaryClass = "hep-ph",
    doi = "10.1103/PhysRevLett.127.052301",
    journal = "Phys. Rev. Lett.",
    volume = "127",
    number = "5",
    pages = "052301",
    year = "2021"
}

@article{Shi:2020htn,
    author = "Shi, Shuzhe and Gale, Charles and Jeon, Sangyong",
    title = "{From chiral kinetic theory to relativistic viscous spin hydrodynamics}",
    eprint = "2008.08618",
    archivePrefix = "arXiv",
    primaryClass = "nucl-th",
    doi = "10.1103/PhysRevC.103.044906",
    journal = "Phys. Rev. C",
    volume = "103",
    number = "4",
    pages = "044906",
    year = "2021"
}

@article{Hu:2021lnx,
    author = "Hu, Jin",
    title = "{Kubo formulae for first-order spin hydrodynamics}",
    eprint = "2101.08440",
    archivePrefix = "arXiv",
    primaryClass = "hep-ph",
    doi = "10.1103/PhysRevD.103.116015",
    journal = "Phys. Rev. D",
    volume = "103",
    number = "11",
    pages = "116015",
    year = "2021"
}

@article{Fu:2021pok,
    author = "Fu, Baochi and Liu, Shuai Y. F. and Pang, Longgang and Song, Huichao and Yin, Yi",
    title = "{Shear-Induced Spin Polarization in Heavy-Ion Collisions}",
    eprint = "2103.10403",
    archivePrefix = "arXiv",
    primaryClass = "hep-ph",
    doi = "10.1103/PhysRevLett.127.142301",
    journal = "Phys. Rev. Lett.",
    volume = "127",
    number = "14",
    pages = "142301",
    year = "2021"
}

@article{Becattini:2021suc,
    author = "Becattini, F. and Buzzegoli, M. and Palermo, A.",
    title = "{Spin-thermal shear coupling in a relativistic fluid}",
    eprint = "2103.10917",
    archivePrefix = "arXiv",
    primaryClass = "nucl-th",
    doi = "10.1016/j.physletb.2021.136519",
    journal = "Phys. Lett. B",
    volume = "820",
    pages = "136519",
    year = "2021"
}

@article{Becattini:2021iol,
    author = "Becattini, F. and Buzzegoli, M. and Inghirami, G. and Karpenko, I. and Palermo, A.",
    title = "{Local Polarization and Isothermal Local Equilibrium in Relativistic Heavy Ion Collisions}",
    eprint = "2103.14621",
    archivePrefix = "arXiv",
    primaryClass = "nucl-th",
    doi = "10.1103/PhysRevLett.127.272302",
    journal = "Phys. Rev. Lett.",
    volume = "127",
    number = "27",
    pages = "272302",
    year = "2021"
}

@article{Hongo:2021ona,
    author = "Hongo, Masaru and Huang, Xu-Guang and Kaminski, Matthias and Stephanov, Mikhail and Yee, Ho-Ung",
    title = "{Relativistic spin hydrodynamics with torsion and linear response theory for spin relaxation}",
    eprint = "2107.14231",
    archivePrefix = "arXiv",
    primaryClass = "hep-th",
    reportNumber = "RIKEN-iTHEMS-Report-21",
    doi = "10.1007/JHEP11(2021)150",
    journal = "JHEP",
    volume = "11",
    pages = "150",
    year = "2021"
}

@article{Weickgenannt:2022zxs,
    author = "Weickgenannt, Nora and Wagner, David and Speranza, Enrico and Rischke, Dirk H.",
    title = "{Relativistic second-order dissipative spin hydrodynamics from the method of moments}",
    eprint = "2203.04766",
    archivePrefix = "arXiv",
    primaryClass = "nucl-th",
    doi = "10.1103/PhysRevD.106.096014",
    journal = "Phys. Rev. D",
    volume = "106",
    number = "9",
    pages = "096014",
    year = "2022"
}

@article{Gallegos:2022jow,
    author = "Gallegos, A. D. and Gursoy, U. and Yarom, A.",
    title = "{Hydrodynamics, spin currents and torsion}",
    eprint = "2203.05044",
    archivePrefix = "arXiv",
    primaryClass = "hep-th",
    doi = "10.1007/JHEP05(2023)139",
    journal = "JHEP",
    volume = "05",
    pages = "139",
    year = "2023"
}

@article{Weickgenannt:2022qvh,
    author = "Weickgenannt, Nora and Wagner, David and Speranza, Enrico and Rischke, Dirk H.",
    title = "{Relativistic dissipative spin hydrodynamics from kinetic theory with a nonlocal collision term}",
    eprint = "2208.01955",
    archivePrefix = "arXiv",
    primaryClass = "nucl-th",
    doi = "10.1103/PhysRevD.106.L091901",
    journal = "Phys. Rev. D",
    volume = "106",
    number = "9",
    pages = "L091901",
    year = "2022"
}

@article{Sarwar:2022yzs,
    author = "Sarwar, Golam and Hasanujjaman, Md and Bhatt, Jitesh R. and Mishra, Hiranmaya and Alam, Jan-e",
    title = "{Causality and stability of relativistic spin hydrodynamics}",
    eprint = "2209.08652",
    archivePrefix = "arXiv",
    primaryClass = "nucl-th",
    doi = "10.1103/PhysRevD.107.054031",
    journal = "Phys. Rev. D",
    volume = "107",
    number = "5",
    pages = "054031",
    year = "2023"
}

@article{Biswas:2022bht,
    author = "Biswas, Rajesh and Daher, Asaad and Das, Arpan and Florkowski, Wojciech and Ryblewski, Radoslaw",
    title = "{Boost invariant spin hydrodynamics within the first order in derivative expansion}",
    eprint = "2211.02934",
    archivePrefix = "arXiv",
    primaryClass = "nucl-th",
    doi = "10.1103/PhysRevD.107.094022",
    journal = "Phys. Rev. D",
    volume = "107",
    number = "9",
    pages = "094022",
    year = "2023"
}

@article{Biswas:2023qsw,
    author = "Biswas, Rajesh and Daher, Asaad and Das, Arpan and Florkowski, Wojciech and Ryblewski, Radoslaw",
    title = "{Relativistic second-order spin hydrodynamics: An entropy-current analysis}",
    eprint = "2304.01009",
    archivePrefix = "arXiv",
    primaryClass = "nucl-th",
    doi = "10.1103/PhysRevD.108.014024",
    journal = "Phys. Rev. D",
    volume = "108",
    number = "1",
    pages = "014024",
    year = "2023"
}

@article{Banerjee:2024xnd,
    author = "Banerjee, Soham and Bhadury, Samapan and Florkowski, Wojciech and Jaiswal, Amaresh and Ryblewski, Radoslaw",
    title = "{Longitudinal spin polarization in a thermal model with dissipative corrections}",
    eprint = "2405.05089",
    archivePrefix = "arXiv",
    primaryClass = "hep-ph",
    doi = "10.1103/923l-yxkc",
    journal = "Phys. Rev. C",
    volume = "111",
    number = "6",
    pages = "064912",
    year = "2025"
}

@article{Wagner:2024fhf,
    author = "Wagner, David and Shokri, Masoud and Rischke, Dirk H.",
    title = "{Damping of spin waves}",
    eprint = "2405.00533",
    archivePrefix = "arXiv",
    primaryClass = "nucl-th",
    doi = "10.1103/PhysRevResearch.6.043103",
    journal = "Phys. Rev. Res.",
    volume = "6",
    number = "4",
    pages = "043103",
    year = "2024"
}

@article{Lin:2024cxo,
    author = "Lin, Shu and Tang, Haiqin",
    title = "{Transient spin modes from relaxational axial kinetic theory}",
    eprint = "2406.17632",
    archivePrefix = "arXiv",
    primaryClass = "nucl-th",
    doi = "10.1103/PhysRevD.110.074042",
    journal = "Phys. Rev. D",
    volume = "110",
    number = "7",
    pages = "074042",
    year = "2024"
}

@article{Bhadury:2024ckc,
    author = "Bhadury, Samapan",
    title = "{Relativistic spin hydrodynamics with momentum- and spin-dependent relaxation time}",
    eprint = "2408.14462",
    archivePrefix = "arXiv",
    primaryClass = "hep-ph",
    doi = "10.1103/PhysRevC.111.034909",
    journal = "Phys. Rev. C",
    volume = "111",
    number = "3",
    pages = "034909",
    year = "2025"
}

@article{Bhadury:2025vvo,
    author = "Bhadury, Samapan and Florkowski, Wojciech and Kar, Sudip Kumar and Mykhaylova, Valeriya",
    title = "{Exact Wigner function for chiral spirals}",
    eprint = "2504.03413",
    archivePrefix = "arXiv",
    primaryClass = "hep-ph",
    doi = "10.1103/vmkx-t59n",
    journal = "Phys. Rev. D",
    volume = "112",
    number = "1",
    pages = "016012",
    year = "2025"
}

@article{Bhadury:2025fil,
    author = "Bhadury, Samapan",
    title = "{Relativistic spin hydrodynamics from novel relaxation time approximation}",
    eprint = "2503.08428",
    archivePrefix = "arXiv",
    primaryClass = "hep-ph",
    doi = "10.1103/hnvm-zfj6",
    journal = "Phys. Rev. C",
    volume = "112",
    number = "2",
    pages = "L021901",
    year = "2025"
}

@article{Buzzegoli:2024mra,
    author = "Buzzegoli, Matteo and Palermo, Andrea",
    title = "{Emergent Canonical Spin Tensor in the Chiral-Symmetric Hot QCD}",
    eprint = "2407.14345",
    archivePrefix = "arXiv",
    primaryClass = "hep-ph",
    doi = "10.1103/PhysRevLett.133.262301",
    journal = "Phys. Rev. Lett.",
    volume = "133",
    number = "26",
    pages = "262301",
    year = "2024"
}

@article{Sapna:2025yss,
    author = "Sapna and Singh, Sushant K. and Wagner, David",
    title = "{Spin polarization of {\ensuremath{\Lambda}} hyperons from dissipative spin hydrodynamics}",
    eprint = "2503.22552",
    archivePrefix = "arXiv",
    primaryClass = "hep-ph",
    doi = "10.1103/1s6g-fs8w",
    journal = "Phys. Rev. C",
    volume = "112",
    number = "5",
    pages = "054902",
    year = "2025"
}

@article{Weickgenannt:2024ibf,
    author = "Weickgenannt, Nora and Blaizot, Jean-Paul",
    title = "{Spin kinetic theory with a nonlocal relaxation time approximation}",
    eprint = "2409.11045",
    archivePrefix = "arXiv",
    primaryClass = "hep-ph",
    doi = "10.1103/PhysRevD.111.056006",
    journal = "Phys. Rev. D",
    volume = "111",
    number = "5",
    pages = "056006",
    year = "2025"
}

@article{Ambrus:2023bid,
    author = "Ambrus, Victor E. and Chernodub, Maxim N.",
    title = "{Rigidly rotating scalar fields: Between real divergence and imaginary fractalization}",
    eprint = "2304.05998",
    archivePrefix = "arXiv",
    primaryClass = "hep-th",
    doi = "10.1103/PhysRevD.108.085016",
    journal = "Phys. Rev. D",
    volume = "108",
    number = "8",
    pages = "085016",
    year = "2023"
}

@article{Patuleanu:2025zbn,
    author = "P{\u{a}}tuleanu, Tudor and Fodor, Amalia Dariana and Ambrus, Victor E. and Crucean, Cosmin",
    title = "{Dirac fermions under imaginary rotation}",
    eprint = "2502.09738",
    archivePrefix = "arXiv",
    primaryClass = "hep-th",
    doi = "10.1103/PhysRevD.111.116004",
    journal = "Phys. Rev. D",
    volume = "111",
    number = "11",
    pages = "116004",
    year = "2025"
}

@article{Ahadi:2025rqs,
    author = "Ahadi, M. Abedlou and Sadooghi, N.",
    title = "{Chiral vortical conductivities and the moment of inertia of a rigidly rotating Fermi gas}",
    eprint = "2502.19264",
    archivePrefix = "arXiv",
    primaryClass = "hep-ph",
    doi = "10.1103/cn47-wdvb",
    journal = "Phys. Rev. D",
    volume = "111",
    number = "11",
    pages = "116011",
    year = "2025"
}

@article{Adler:1976zt,
    author = "Adler, Stephen L. and Collins, John C. and Duncan, Anthony",
    title = "{Energy-Momentum-Tensor Trace Anomaly in Spin 1/2 Quantum Electrodynamics}",
    reportNumber = "COO-2220-77-REV, COO-2220-77",
    doi = "10.1103/PhysRevD.15.1712",
    journal = "Phys. Rev. D",
    volume = "15",
    pages = "1712",
    year = "1977"
}

@article{Collins:1976yq,
    author = "Collins, John C. and Duncan, Anthony and Joglekar, Satish D.",
    title = "{Trace and Dilatation Anomalies in Gauge Theories}",
    reportNumber = "COO-2220-88",
    doi = "10.1103/PhysRevD.16.438",
    journal = "Phys. Rev. D",
    volume = "16",
    pages = "438--449",
    year = "1977"
}

@article{Das:2025kgq,
    author = "Das, Aritra and Tuchin, Kirill",
    title = "{Rotational stability of magnetic field in rotating quark-gluon plasma}",
    eprint = "2502.18354",
    archivePrefix = "arXiv",
    primaryClass = "hep-ph",
    doi = "10.1016/j.nuclphysa.2025.123075",
    journal = "Nucl. Phys. A",
    volume = "1059",
    pages = "123075",
    year = "2025"
}
